\documentclass[fleqn,usenatbib]{mnras}

\usepackage{newtxtext,newtxmath}

\usepackage[T1]{fontenc}

\DeclareRobustCommand{\VAN}[3]{#2}
\let\VANthebibliography\thebibliography
\def\thebibliography{\DeclareRobustCommand{\VAN}[3]{##3}\VANthebibliography}

\usepackage{graphicx}	
\usepackage{amsmath}  
\usepackage{graphicx}
\usepackage{subcaption}
\usepackage[export]{adjustbox}
\usepackage{wrapfig}
\usepackage{multicol}
\usepackage{lipsum}
\usepackage{tikz}
\usepackage{float}

\title[A statistical analysis of NGC 6416]{Bayesian and Statistical Analysis of the Open Star Cluster NGC 6416}

\author[Deka . S] {
Simanta Deka$^{1}$\thanks{E-mail: simanta1955@gmail.com}
\\
$^{1}$Department of Physics, Gauhati University, Guwahati 781014, Assam, India\\
}

\date{Accepted XXX. Received YYY; in original form ZZZ}

\pubyear{\the\year{}}

\begin{document}
\label{firstpage}
\pagerange{\pageref{firstpage}--\pageref{lastpage}}
\maketitle

\begin{abstract}
In our Bayesian and Statistical Analysis investigation of the open cluster NGC 6416, we utilized Gaia EDR3 astrometry data and ensemble-based unsupervised machine learning techniques to identify 406 cluster members. Using the MESA Isochrones and Stellar Tracks (MIST) with Gaia EDR3 data, we determined the following parameters for NGC 6416: a distance of approximately 1021
pc, an age of about $12.58\pm 0.1$ Myr, a metallicity (z) of roughly $0.032\pm0.0015$, a binarity fraction near $0.419\pm0.021$, and an extinction ($A_V$) of approximately $0.995\pm0.058$ mag for an $R_V$ value of around $3.064\pm0.102$. We also fitted the radial surface density profile and conducted orbit analysis of the cluster using galpy. And finally found that the star formation scenarios are not observed in the open star cluster NGC 6416.
\end{abstract}




\section{Introduction}

Open star clusters are groups of stars that share similar ages, chemical compositions, and kinematics, as they originate from the same molecular cloud. However, distinguishing cluster members from the numerous field stars is challenging, as most open clusters are situated on the densely populated galactic plane (\citet{castro}). The accurate identification of cluster members is crucial for studying the characteristics of clusters and understanding star formation. We placed considerable emphasis on accurate astrometric data, particularly parallax and proper motion measurements, to differentiate clustered stars from field stars. This can be determined as stars that are at a similar distance from the cluster by utilizing parallax, which provides precise information about each star's distance (\citet{kharchenko}). Because cluster members typically share a common velocity owing to their shared origin, proper motion data, which track the apparent movement of stars across the sky, helps identify stars that move in a similar direction and at comparable rates. The method has been completely transformed by the advent of high-precision satellite missions, such as Gaia, which offer remarkably accurate measurements of parallax and proper motion for millions of stars. Utilizing statistical and machine learning techniques on these data enables a much more reliable distinction between cluster members and the field population than was previously possible (\citet{(Debetal.2022)}). Studying open clusters yields crucial insights into star formation along with the technical challenges of identifying their members. By examining the distribution, ages, and masses of stars within these clusters, a deeper understanding of the initial conditions of star formation, the efficiency of gas conversion into stars, and the dynamical evolution of a cluster can be found (\citet{2003ARA&A}). A deep understanding of the intricate processes involved in star formation, stellar evolution, and galactic dynamics relies on the study of the early open-star clusters. Typically a few million to a few hundred million years old, these clusters provide a unique perspective on the early stages of stellar and planetary system development, along with insights into the conditions and processes that drive star formation.

NGC 6416 is an open cluster in the constellation Scorpius. NGC 6416 is located south of the celestial equator, hence it is better visible from the southern hemisphere. It is centrally located at an equatorial coordinates (RA($\alpha$)= $266.014$, Dec($\delta$)= -$32.344$) (\citet{Tarricq.Y.2021}) and its galactic longitude (l) = $356.9179^{\circ}$, galactic latitude (b) = -$01.4930^{\circ}$. From the UBV photographic photometry of stars in this region, 20 stars belonging to NGC 6416 have been measured (\citet{1972}). The most prominent members among the measured stars are those of spectral type $B_8$, which belong to the B-class. These stars are very hot and luminous. An $F_0$ star, classified as an F-type star, is moderately hot and appears white to yellow-white in color, while an $F_8$ star is also present in NGC 6416 which is cooler than $F_0$ stars. A $G_2$ star is part of the G-class, which typically consists of yellow stars. The measured stars, which are projected into the neighbourhood of NGC6416, indicate that the real level of absorption is in the distance range of 0.5 to 1,6 kpc (\citet{1972}). Inspection of the Henry Draper Catalogue shows that the earliest spectral type of the stars in the cluster area is $B_8$. NGC6416 is a cluster of intermediate age, the upper part of the main sequence is influenced by evolutionary effects (\citet{1970}). One of the most significant chemical and age indicators of Milky Way development is open clusters. Although they provide crucial constraints on galaxy evolution, the use of open clusters has been limited due to discrepancies in sufficient measurements from various studies (\citet{Ray_2022}). The selected measurements for NGC 6416 are provided as follows log(age)= 8.36 yrs, [Fe/H] = -$0.07\pm0.11$, [Si/Fe] = $0.02 \pm 0.04$, [Al/Fe] = $0.03 \pm 0.07$, [Mg/Fe] = -$0.03 \pm 0.04$, [O/Fe] = $0.01 \pm 0.03$ (\citet{Ray_2022}).

The structure of this work is given as follows. In section \ref{archive} we described the source of data for open star cluster NGC 6416, taking data from the data bases such as Gaia DR2 and Gaia EDR3 (\citet{Gaia2018};\citet{Gaia2021}). In section \ref{member} NGC 6416's cluster membership is determined using ensemble-based unsupervised machine learning techniques (\citet{(Debetal.2022)}). In section \ref{structure} the radial surface density profile of NGC 6416 is fitted using a King profile \citet{KING(1962)}. In section \ref{isochrone} we calculate various isochrone properties of NGC 6416 using MESA Isochrones and
Stellar Tracks (MIST) such as distance, metallicity, age, extinction or reddening, binary fraction, and the total-to-selective extinction ratio. In section \ref{orbital} we determine various orbital parameters such as birth radii and galactic distribution using galpy a software package for galactic dynamics (\citet{Sheikh.A.H&MedhiBimanJ2024}). In section \ref{results} we presented the results and compared them with previous studies for enhanced reference and accuracy, and subsequently, we summarized and discussed various other properties of NGC 6416 in section \ref{summary}.

\section{ARCHIVAL DATA SETS}
\label{archive}
\subsection{ Gaia data}
We utilised Gaia Early Data Release 3 (Gaia EDR3; \citet{2016A&A...595A...1G}) for doing the cluster membership analysis in our study. The initial installment of the third Gaia data release, Gaia EDR3, includes astrometry and photometry, along with radial velocities taken from Gaia DR2 after the removal of a few spurious entries. It comprises data on the position of the sky, parallax, and proper motions for about 1.468 billion sources, that have a limiting magnitude of G=21 and a bright limit of G=3. The Gaia EDR3 catalogue is mostly complete between G=12 and G=17 (\citet{Gaia2021}).

It covers astrometric parameters as the position uncertainties are 0.01-0.03 mas for G<15, 0.05 mas for G=17, 0.5 mas at G=20, and 1.0 mas at G=21 mag. Similarly, for parallax uncertainties, it ranges from 0.02-0.03 mas for G<15 to 1.3 mas at G=21 mag, and for proper motion uncertainties, it goes from 0.02-0.03 mas for G<15 to 1.4 $masyr^{-1}$ at G=21 mag. The catalouge also produces photometric data based on 3 bands: G, $G_{BP}$, $G_{RP}$. The uncertainty values range from 0.3 mmag for G<13 to 6 mmag at G=20 mag for the G-band, and the radial velocities from Gaia DR2 have been added to Gaia EDR3; it contains median radial velocities for about 7.21 million stars with a mean G magnitude between 4 and 13.

We utilised Gaia Data Release 2 (Gaia DR2;\citet{Gaia2018}) for estimating distance from parallax. The Gaia DR2
provides us with a five-parameter astrometric solution: positions,
parallaxes ($\pi$), and proper motions ($\mu_\alpha \cos \delta$, $\mu_\delta$ ) for over 1.3 billion
sources.

\section{MEMBERSHIP ANALYSIS}
\label{member}
Using ensemble-based unsupervised machine learning approaches, we have implemented an enhanced approach for estimating the membership probability of the open cluster NGC 6416. This method is highly effective in distinguishing the cluster members from the field stars (\citet{(Debetal.2022)}). The first step involves choosing an appropriate range of parameters $(\pi, \mu_\alpha \cos \delta, \mu_\delta)$ for the membership analysis. In order to carry out this step, a smaller sample of cluster (as well as field star) data is downloaded from the Gaia EDR3 data repository within a very limited search radius; in our case, we took about 5 arcmin of search radius using the technique called the k-nearest neighbor (kNN) algorithm. In the second step of the cluster membership analysis, a Gaussian Mixture Model (GMM) is used to determine the Mahalanobis distance (MD) based on three astrometric parameters $(\pi, \mu_\alpha \cos \delta, \mu_\delta)$ within a wider search radius of 90 arcmin. This reduces the dimension from 3D to 1D by eliminating field stars, making it easier to identify cluster members (\citet{(Debetal.2022)}).

A combination of two Gaussian distributions, one for the cluster stars and the other for the field stars, can be used to depict the distribution of Mahalanobis Distances (MD) in one dimension. A two-component Gaussian Mixture Model (GMM) is used to examine this one-dimensional Gaussian distribution. The expectation-maximization (EM) approach is used to fit the model and provides information on the stars' membership probabilities in the open cluster (\citet{(Debetal.2022)}).

\subsection{kNN Technique for Outlier Removal}
The k-Nearest Neighbors (kNN) algorithm (\citet{CoverandHart1967}) is used to find and eliminate likely outliers. To eliminate outliers from the sample, which are often recognized as field stars, a threshold that is determined by the average nearest neighbor distance $(\overline{d}_{\text{NN}})$ is used. The Euclidean distances between each star and its closest neighbors within the three-dimensional parameter space described by $(\pi, \mu_\alpha \cos \delta, \mu_\delta)$ are averaged to determine the $\overline{d}_{\text{NN}}$. Outlier removal is the process of highlighting stars that are significantly apart from their neighbors, indicating possible outliers (\citet{(Debetal.2022)}). The average nearest neighbor distance, $\overline{d}_{\text{NN}}$, is expressed as follows:
\begin{equation}
\overline{d}_{\text{NN}} = \sum_{k} \frac{d(x,k)}{\text{NN}_k} \tag{1},
\end{equation}
In this equation, $\text{d}(x,k)$ represents the distance between an instance x and its k nearest neighbors, where $\text{NN}_k$ represents the entire number of nearest neighbors. In general, the cluster star's nearest neighbor distance should be smaller than that of the field stars (\citet{(Debetal.2022)}). 

The purpose of the cluster membership process is to eliminate these outliers, especially field stars. A skewed estimate of membership probability can arise from the cluster star distribution being distorted by an excessive number of outliers in the sample. Smaller cutoff values can be selected to effectively minimize outliers while preserving a large number of cluster stars for correct analysis by looking at the distribution of average nearest neighbor distances. The distance is calculated using the Euclidean metric. The Euclidean distance $\text{D}_E(x,y)$ between two points x and y in an n-dimensional space is defined by the sum of the squared differences between corresponding components of the points (\citet{(Debetal.2022)}) and it is expressed as, 

\begin{equation}
\text{D}_E(x, y) = \sqrt{\sum_{i=1}^{n} (x_i - y_i)^2}\ \tag{2},
\end{equation}
where n is the number of dimensions, and $\text{x}_i$ and $\text{y}_i$ represent the $\text{i}_th$ components of x and y, respectively.

\begin{figure*}
    \centering
    \begin{subfigure}{0.33\textwidth}  
        \centering
        \includegraphics[width=\linewidth]{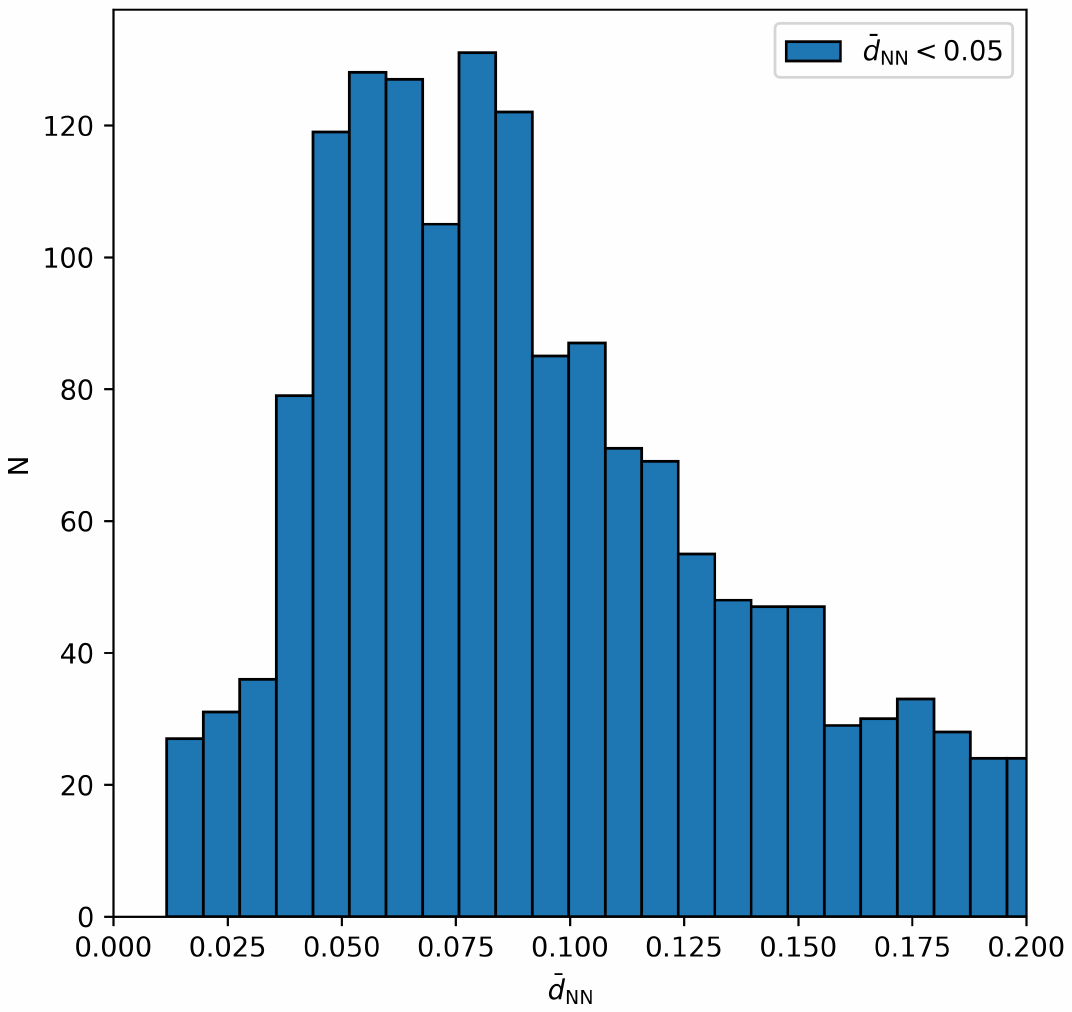}
        \caption{}
        \label{fig:image1}
    \end{subfigure}
    \hfill
    \begin{subfigure}{0.33\textwidth}  
        \centering
        \includegraphics[width=\linewidth]{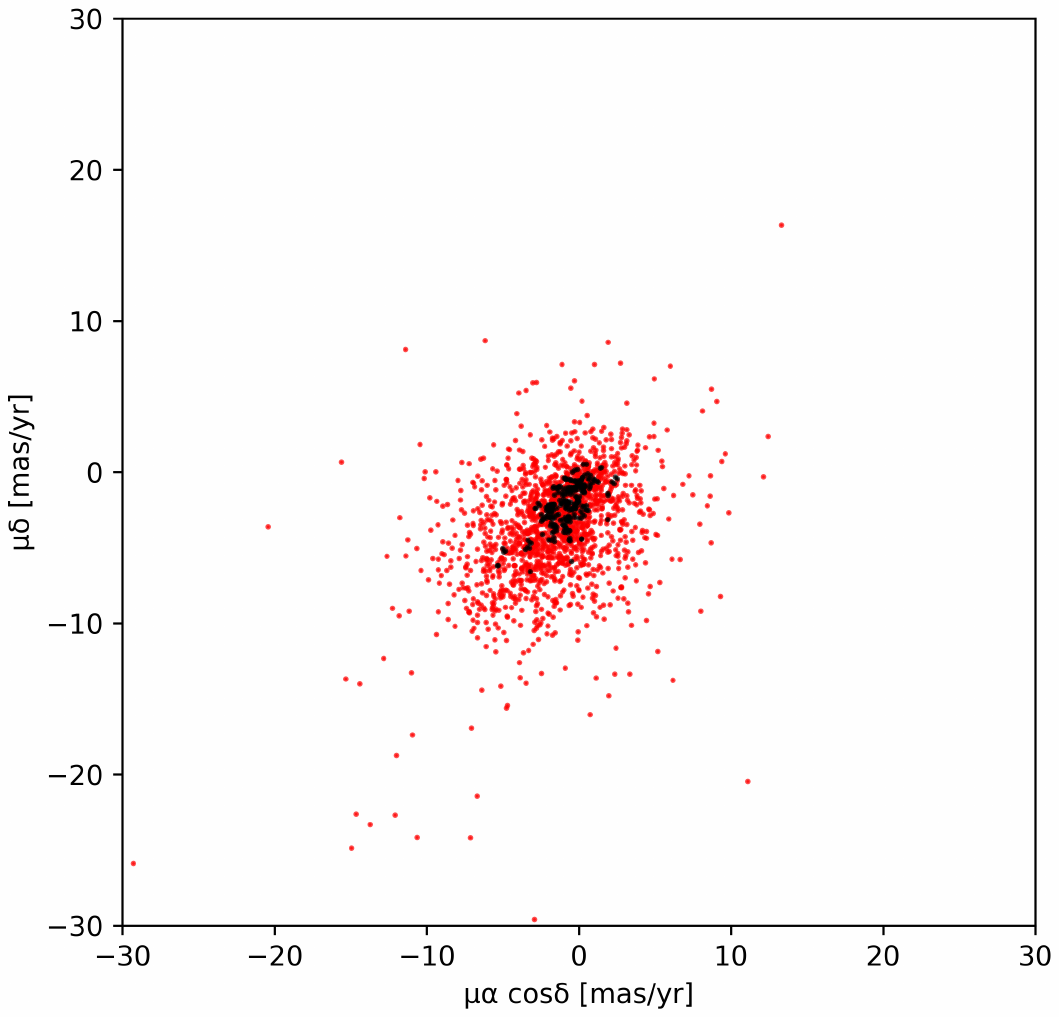}
        \caption{}
        \label{fig:image2}
    \end{subfigure}
    \hfill
    \begin{subfigure}{0.33\textwidth}  
        \centering
        \includegraphics[width=\linewidth]{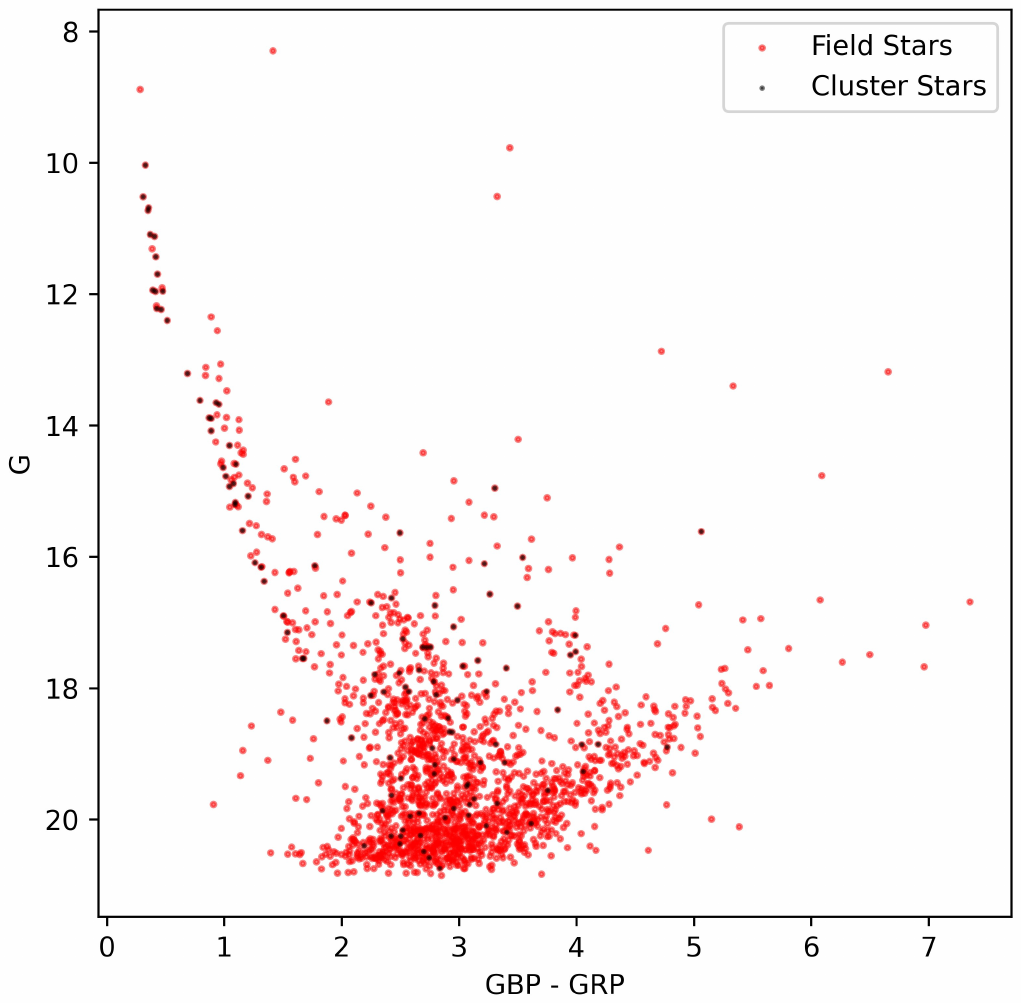}
        \caption{}
        \label{fig:image3}
    \end{subfigure}
    
    \caption{(a) Nearest Neigbour distribution of 1838 stars within a 5 arcmin explore radius for the NGC 6416 cluster; (b) Cluster stars (in black) with a proper motion distribution selected with $\overline{d}_{\text{NN}}$ = 0.05 across 1838 stars are plotted; (c) Diagram of the color-magnitude of the selected cluster stars. }
    \label{fig:three_images_side_by_side1}
\end{figure*}

\begin{figure*}
    \centering
    \begin{subfigure}{0.33\textwidth}  
        \centering
        \includegraphics[width=\linewidth]{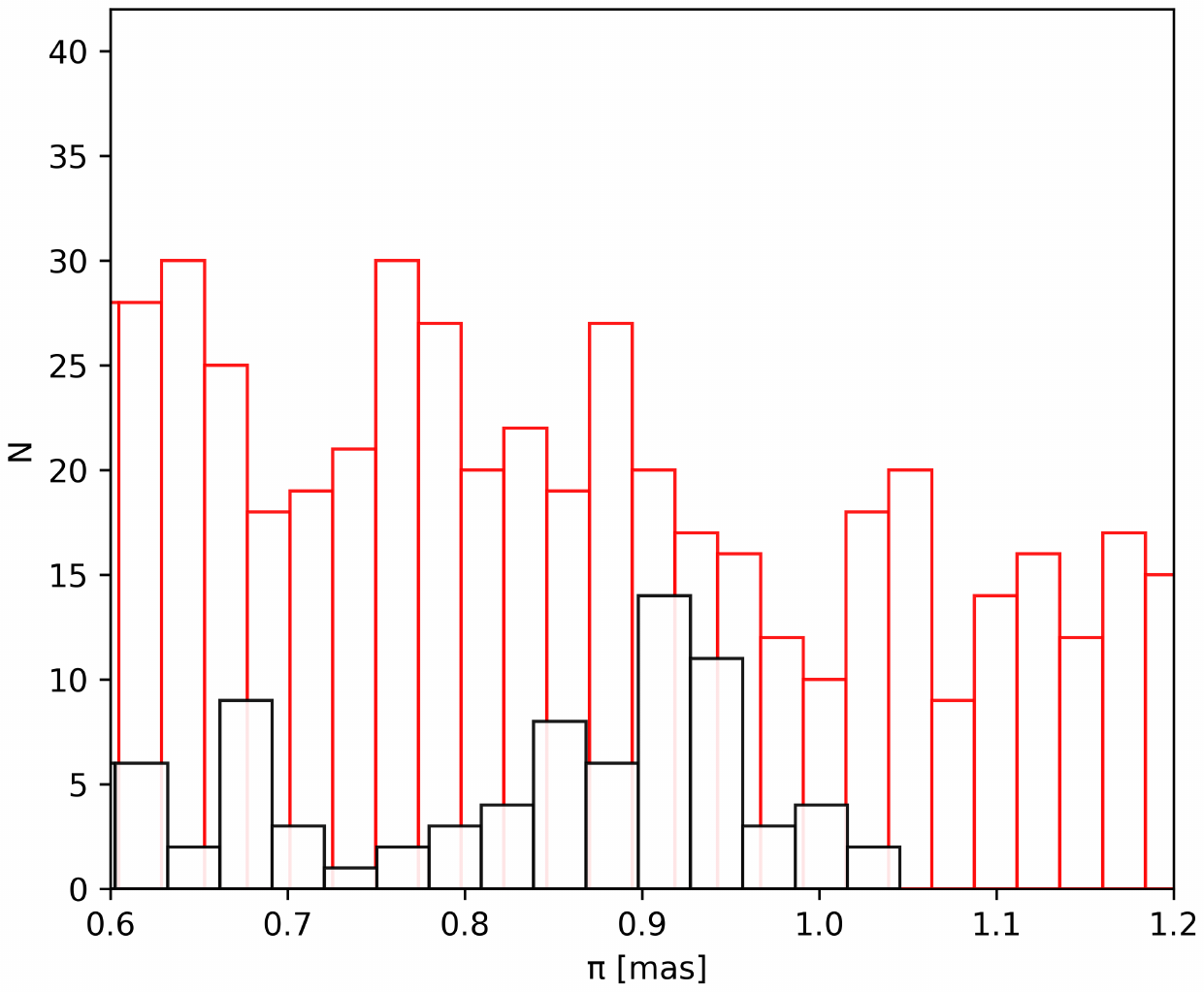}
        \caption{}
        \label{fig:image4}
    \end{subfigure}
    \hfill
    \begin{subfigure}{0.33\textwidth}  
        \centering
        \includegraphics[width=\linewidth]{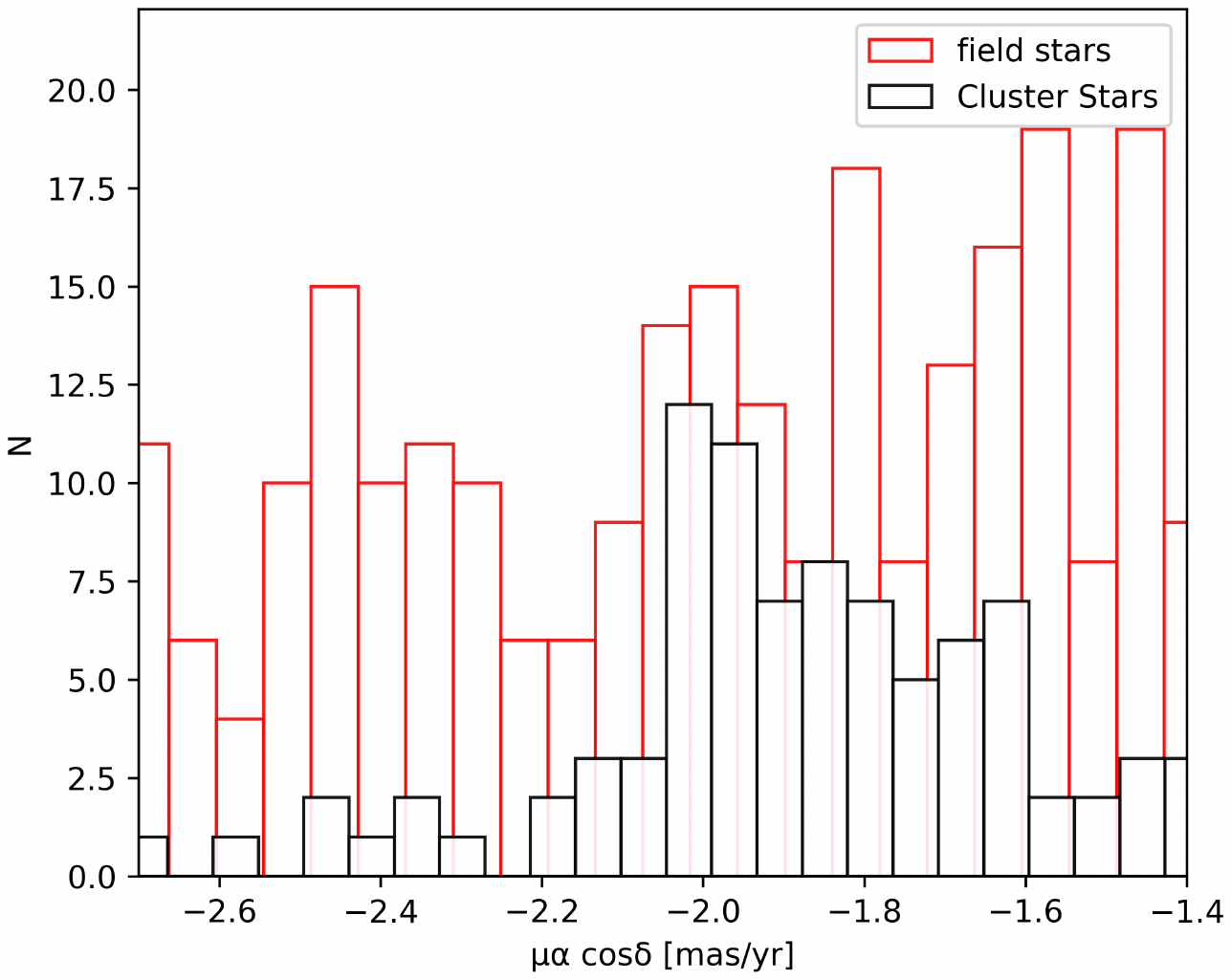}
        \caption{}
        \label{fig:image5}
    \end{subfigure}
    \hfill
    \begin{subfigure}{0.33\textwidth}  
        \centering
        \includegraphics[width=\linewidth]{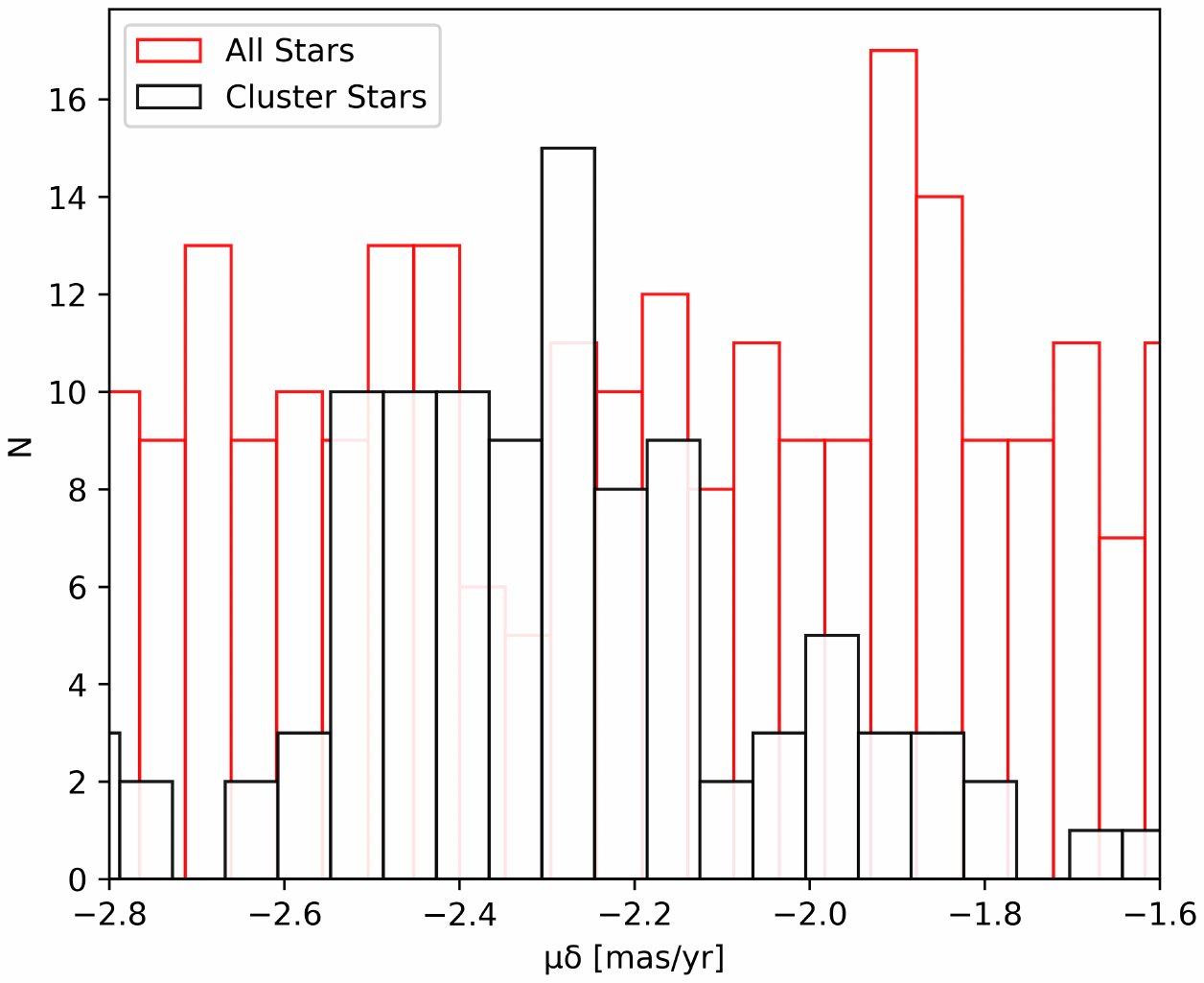}
        \caption{}
        \label{fig:image6}
    \end{subfigure}
    
    \caption{The selected cluster stars (orange) show their histogram plots for $\pi$, $\mu_\alpha \cos \delta$, $\mu_\delta$ in (a), (b), and (c) figures, respectively. }
    \label{fig:three_images_side_by_side2}
\end{figure*}

\begin{table}
    \caption{Parameter $(\pi, \mu_\alpha \cos \delta, \mu_\delta)$ range for cluster NGC 6416 }
    \begin{tabular}{|c|c|c|c|}
        \hline
        $\overline{d}_{\text{NN}}$ & $\pi(mas)$ & $\mu_\alpha \cos \delta(mas/yr)$ & $\mu_\delta(mas/yr)$ \\ \hline
        0.05 & [0.7, 1.1] & [-2.6, -1.5] & [-2.7, -1.7] \\ \hline
    \end{tabular}
    \label{table1}
\end{table}

This procedure is demonstrated using parallax and proper motion data for 19438 stars within a 5 arcminute search radius obtained from the Gaia EDR3 database. The search criteria comprised stars with a parallax of at least zero. The nearest neighbor distance $(\overline{d}_{\text{NN}})$ for each star was determined using a three-dimensional parameter space $(\pi, \mu_\alpha \cos \delta, \mu_\delta)$. A cut-off threshold of $\overline{d}_{\text{NN}}$= 0.05 was used to remove outliers, resulting in a sample of 1838 stars. The surviving stars were then plotted into the appropriate motion in fig 1(b) and color-magnitude diagrams in fig 1(c) (\citet{(Debetal.2022)}). It is worth mentioning that within a limited search radius, most cluster stars are often located in a densely populated location (\citet{(Debetal.2022)}). The panels of Fig. 2 show histograms of the distribution of 1838 stars in the $\pi$, $\mu_\alpha \cos \delta$ and $\mu_\delta$ spaces. The strong peaks in these parameter distributions make it simple to differentiate between cluster stars. These distributions allow for the easy identification of appropriate parallax and motion ranges. The list of parameters and their respective comparisons are shown in the table \ref{table1}. A bigger search radius of 90 arcmin with parallax \( \geq \) 0 produces 1048576 stars. Using the previously established parameter range, the sample is reduced to 3389 stars for further investigation (\citet{(Debetal.2022)}).

\subsection{Gaussian Mixture Model on MD Distributions}

Mahalanobis distance (\citet{Mahalanobis1927}) is a robust multi-variate distance metric which measures the distance between a point and a distribution. It is a multidimensional generalization of the idea of measuring how many standard deviations away from the point is the mean of the distribution (\citet{(Debetal.2022)}). The MD of an observation $\vec{x}=(x_1,x_2,....,x_n)$ from a set of observations with mean $\vec{\mu}=(\mu_1,\mu_2,....,\mu_n)$ and covariance matrix  is defined as (\citet{Mahalanobis1927}),

\begin{equation}
D_M(\vec{x}) = \sqrt{(\vec{x} - \vec{\mu})^\top \Sigma^{-1} (\vec{x} - \vec{\mu})},
 \tag{3}
\end{equation}

\begin{figure*}
    \centering
    \begin{subfigure}{0.33\textwidth}  
        \centering
        \includegraphics[width=\linewidth]{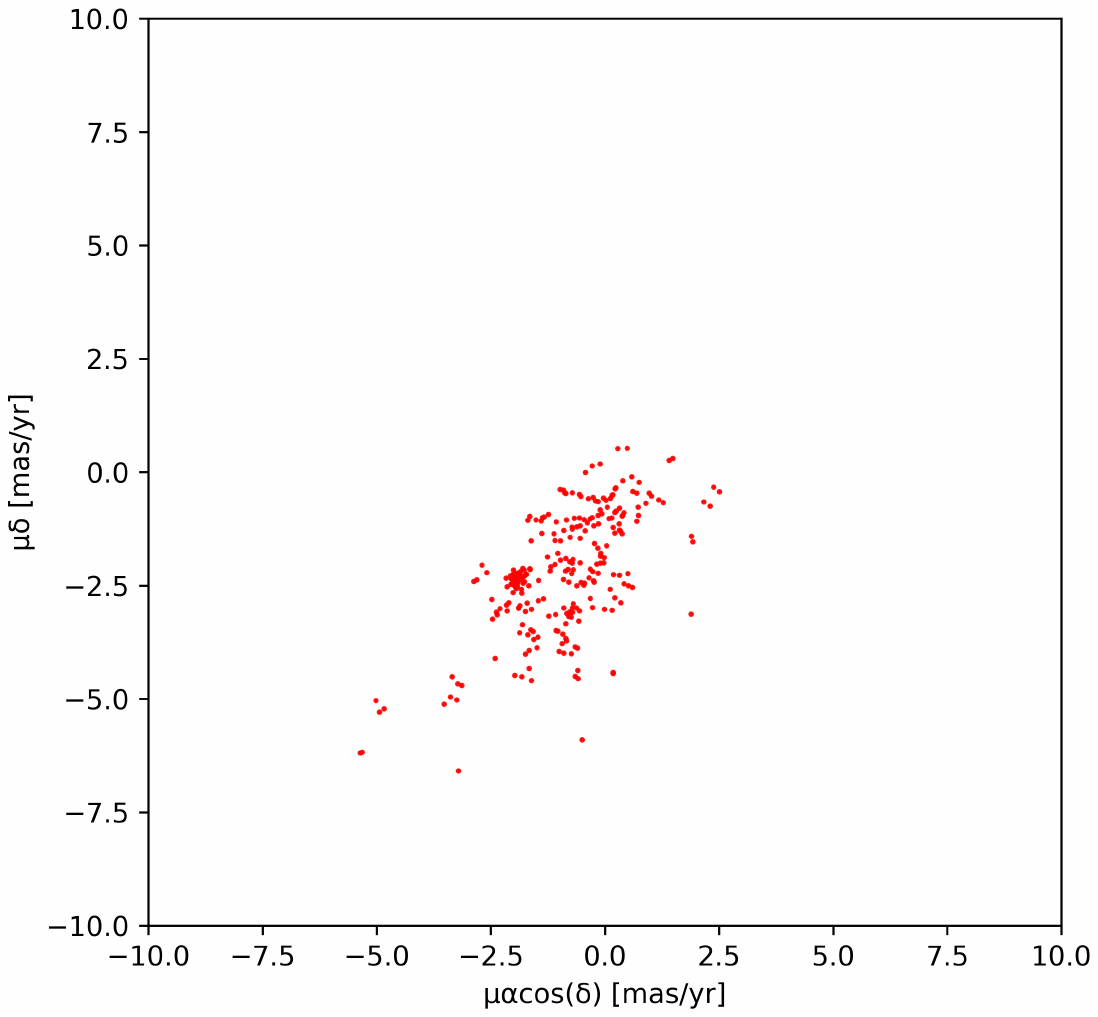}
        \caption{}
        \label{fig:image7}
    \end{subfigure}
    \hspace{-0.5cm}
    \begin{subfigure}{0.33\textwidth} 
        \centering
        \includegraphics[width=\linewidth]{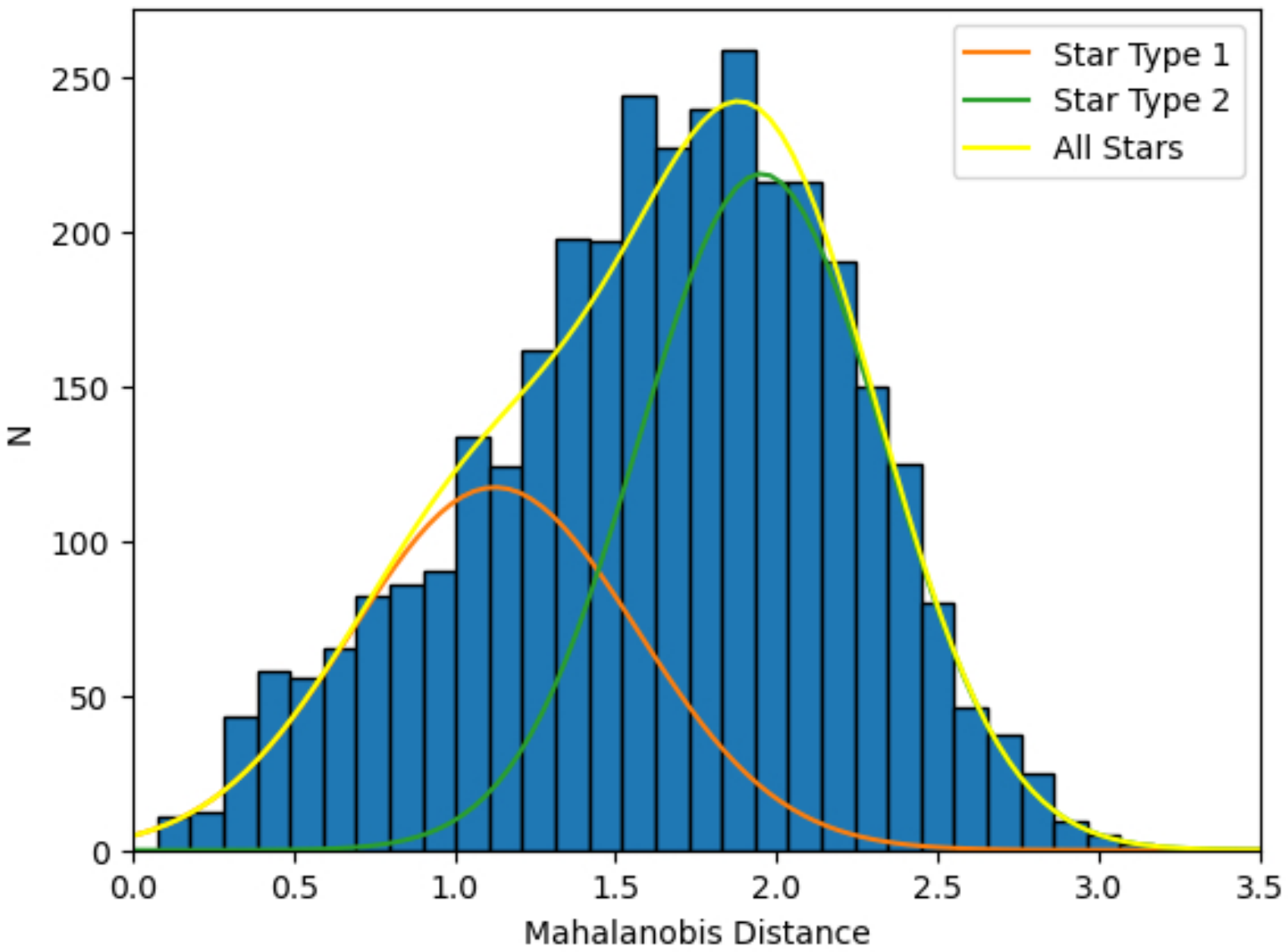}
        \caption{}
        \label{fig:image8}
    \end{subfigure}
    \caption{(a)The proper motion plot of the NGC 6416 cluster, featuring 3389 stars, was generated using a selected set of parameters from 1048576 stars with $parallax\geq 0$, and was downloaded within a search radius of 90 arcminutes;(b) The two-component GMM model of the MD distribution for these 3389 stars and the resulting fits for cluster and field stars, as well as both combined.}
    \label{fig:three_images_side_by_side3}
\end{figure*}

\begin{figure*}
    \centering
    \begin{subfigure}{0.33\textwidth}  
        \centering
        \includegraphics[width=\linewidth]{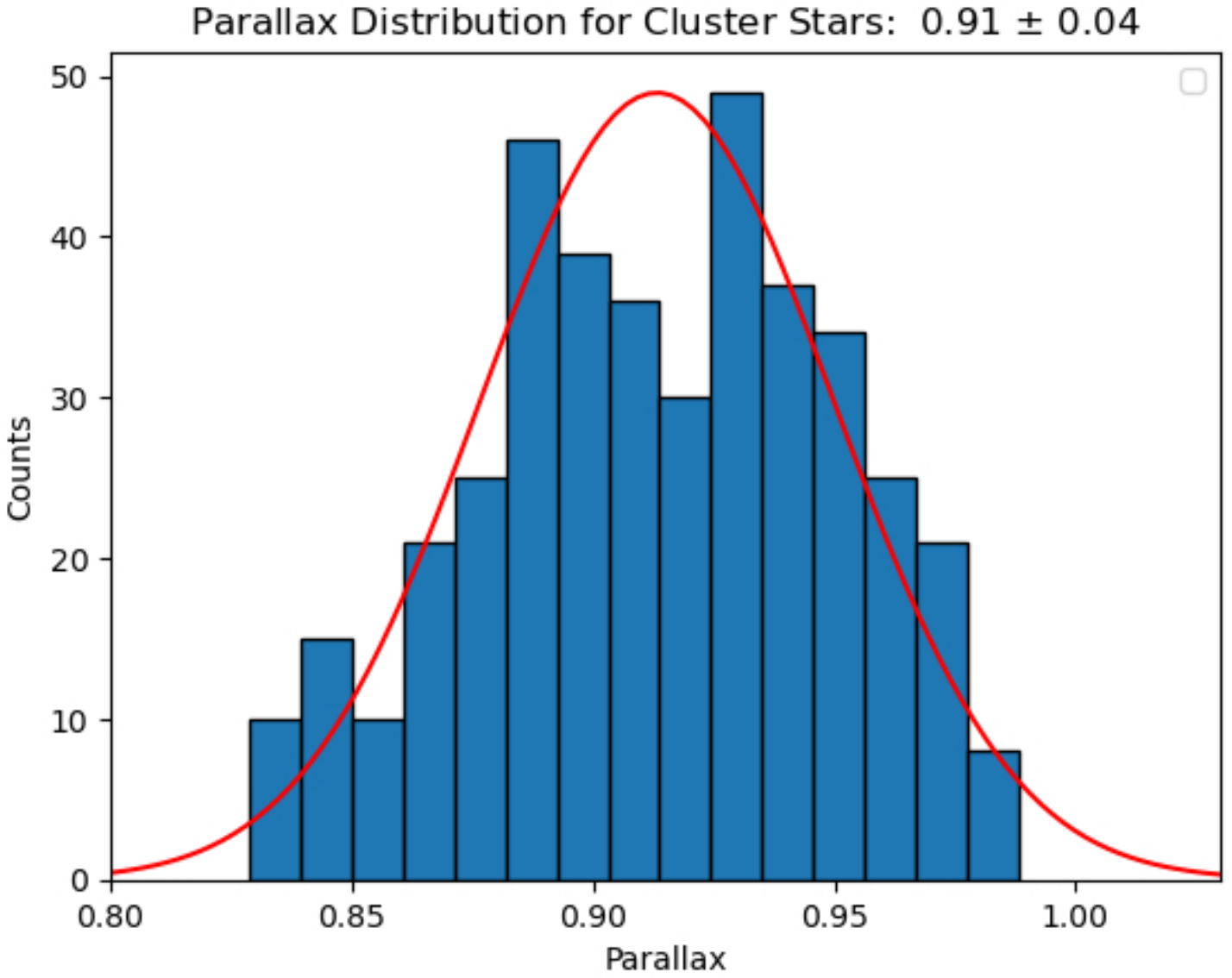}
        \caption{}
        \label{fig:image9}
    \end{subfigure}
    \hfill
    \begin{subfigure}{0.33\textwidth}  
        \centering
        \includegraphics[width=\linewidth]{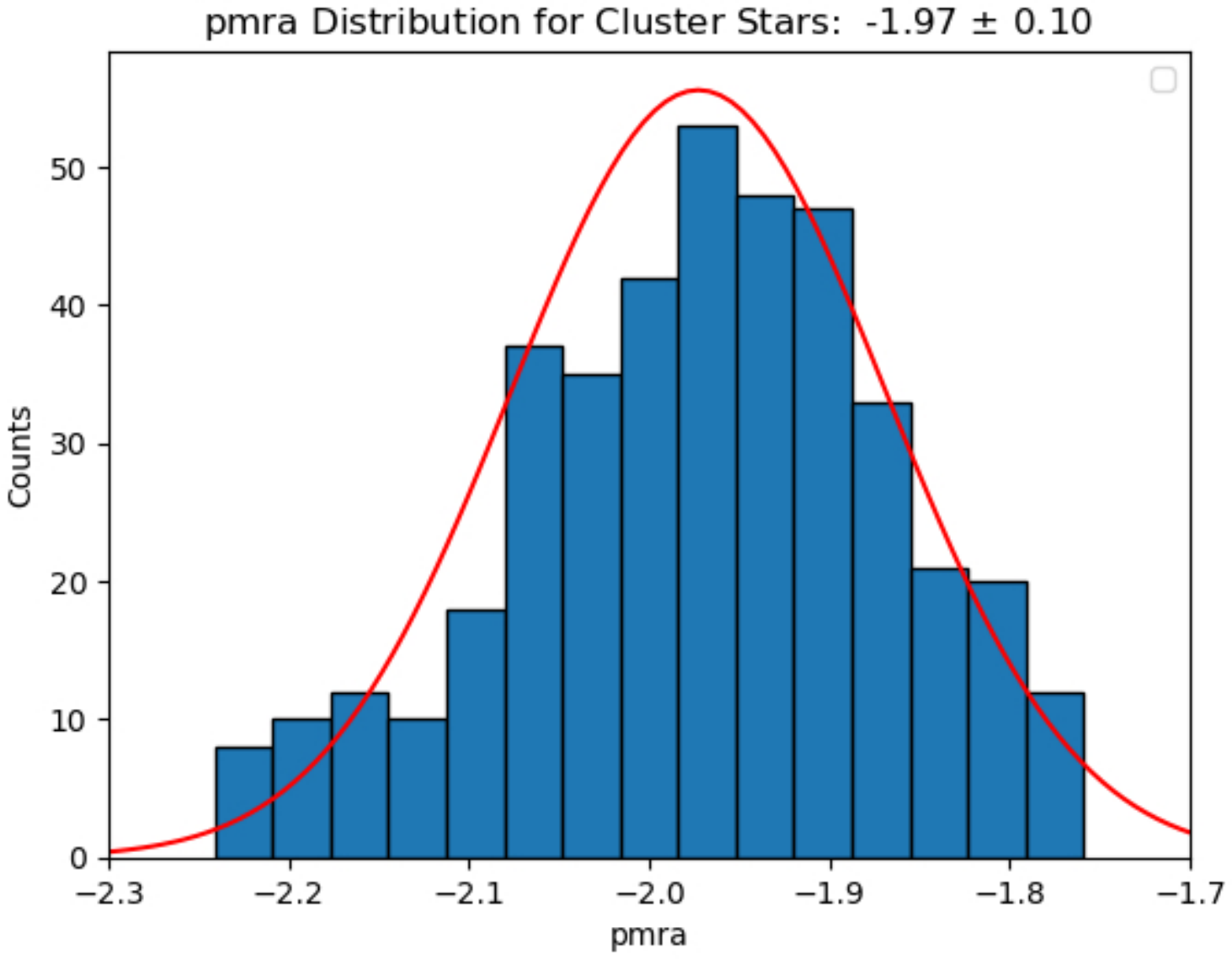}
        \caption{}
        \label{fig:image10}
    \end{subfigure}
    \hfill
    \begin{subfigure}{0.33\textwidth}  
        \centering
        \includegraphics[width=\linewidth]{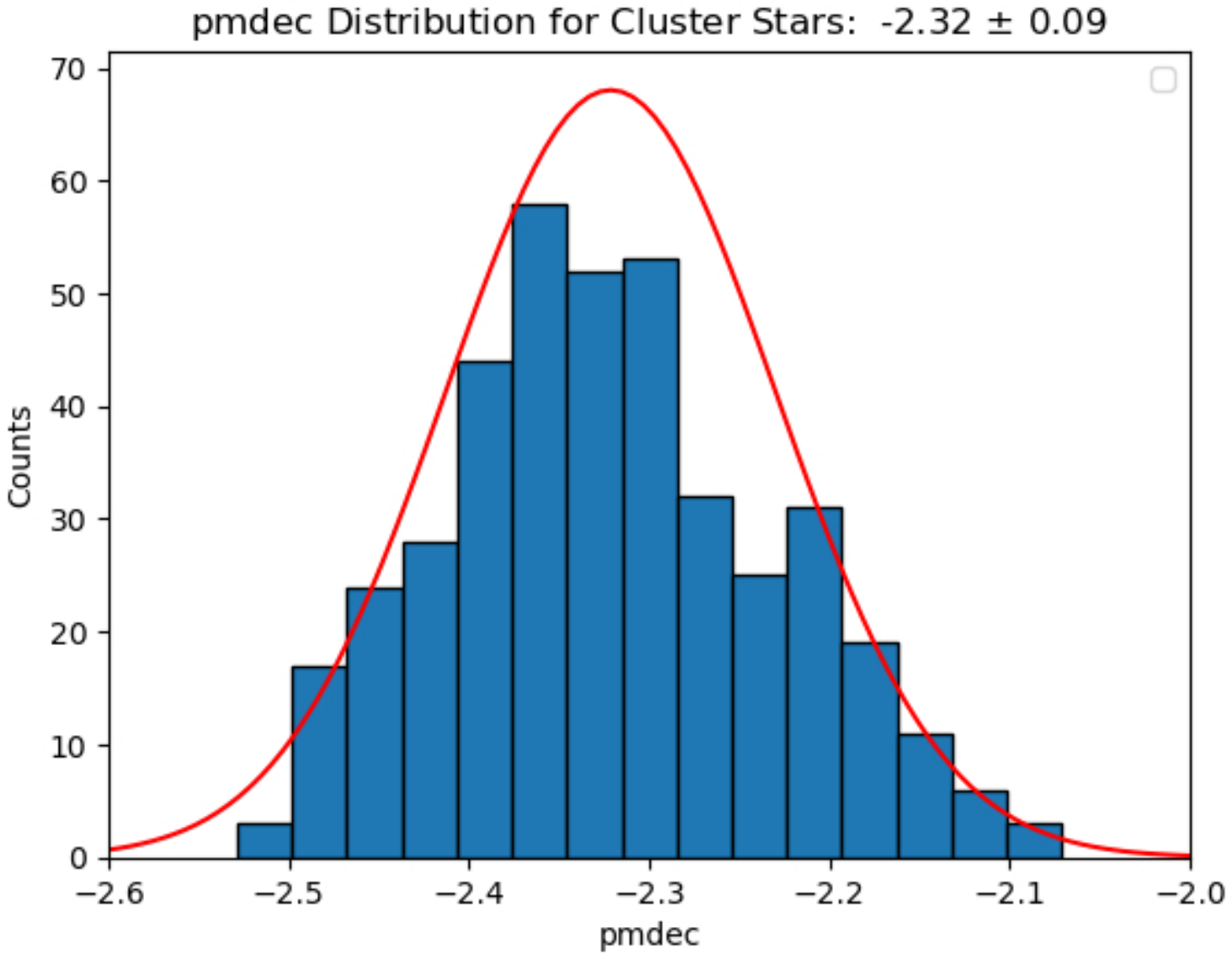}
        \caption{}
        \label{fig:image11}
    \end{subfigure}
    
    \caption{Gaussian fits are employed to model the parallax and individual proper motion distributions, allowing us to derive the cluster mean values for stars with a membership probability exceeding 0.50.}
    \label{fig:three_images_side_by_side4}
\end{figure*}

\begin{figure*}
    \centering
    \begin{subfigure}{0.33\textwidth}  
        \centering
        \includegraphics[width=\linewidth]{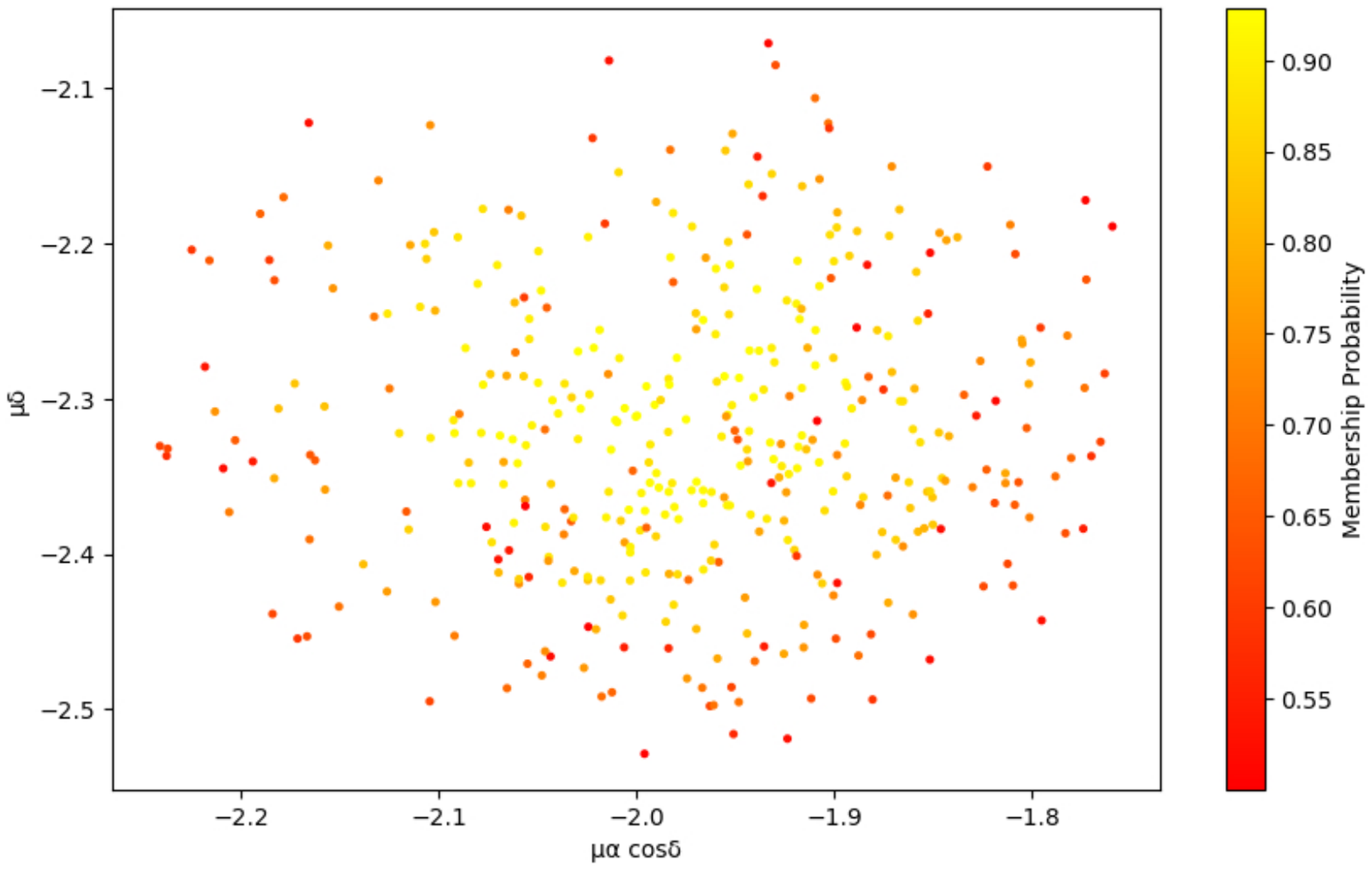}
        \caption{}
        \label{fig:image12}
    \end{subfigure}
    \hfill
    \begin{subfigure}{0.33\textwidth}  
        \centering
        \includegraphics[width=\linewidth]{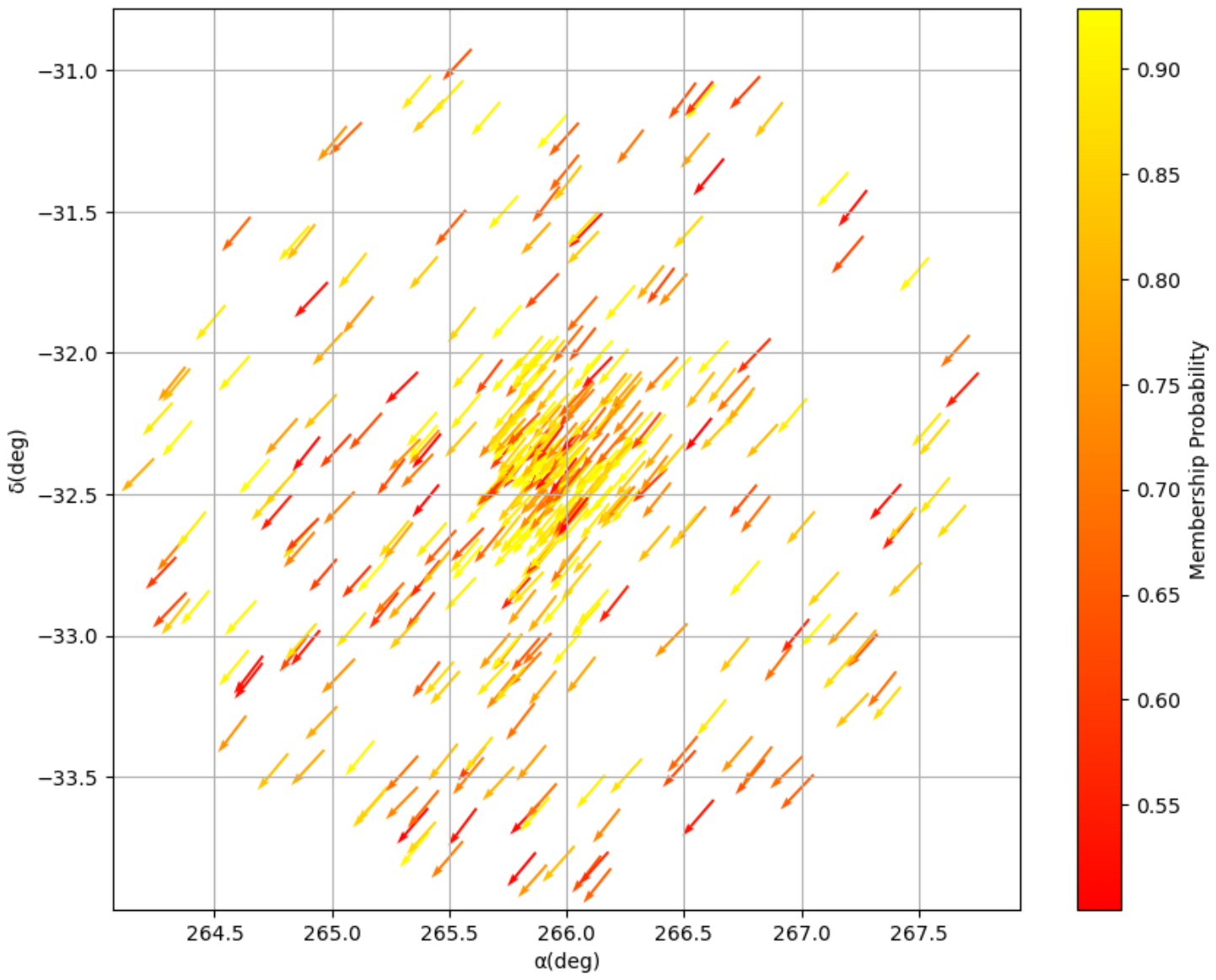}
        \caption{}
        \label{fig:image13}
    \end{subfigure}
    \hfill
    \begin{subfigure}{0.33\textwidth}  
        \centering
        \includegraphics[width=\linewidth]{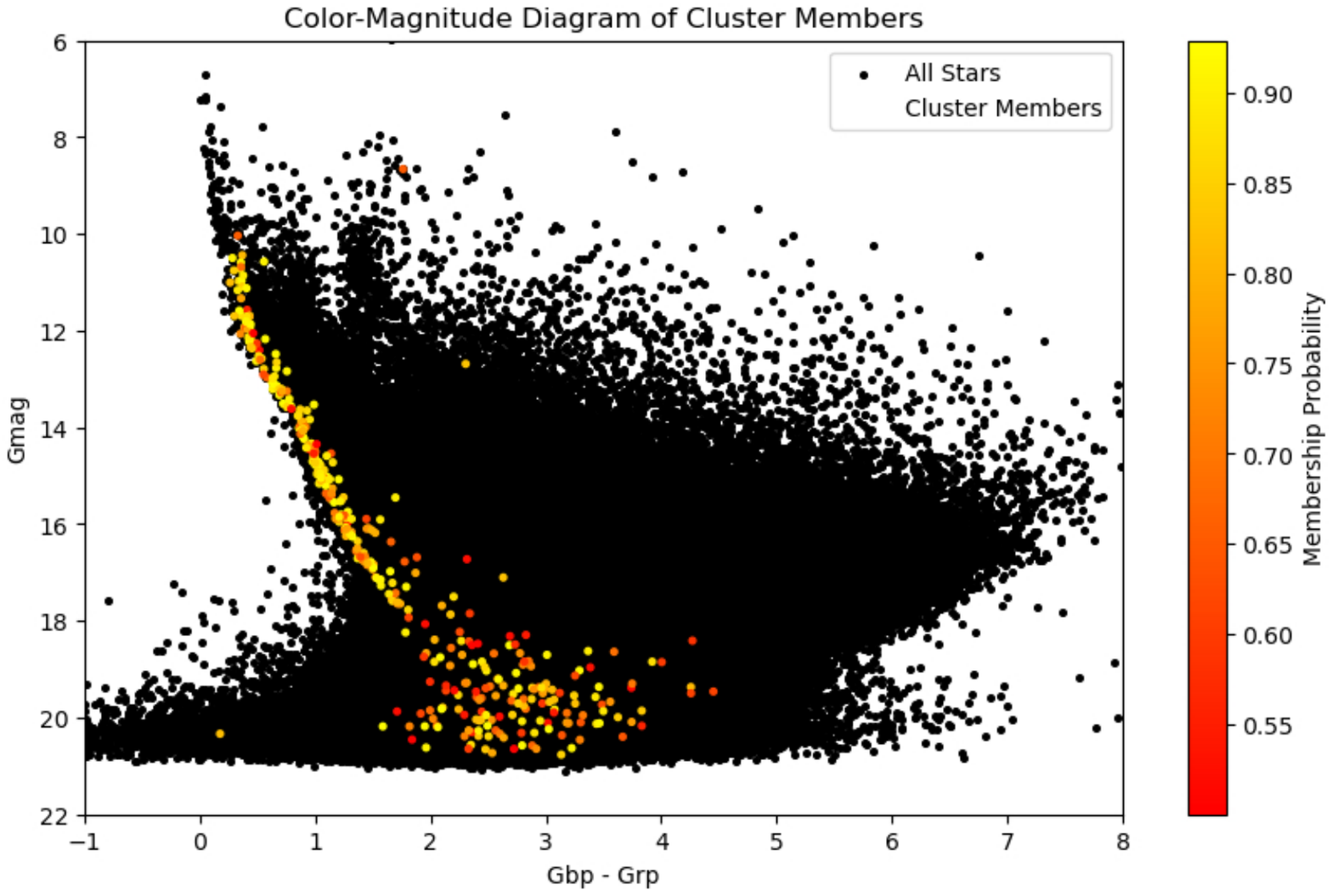}
        \caption{}
        \label{fig:image14}
    \end{subfigure}
    
    \caption{(a) The member stars of the cluster are color-coded in the proper motion plane according to the likelihood that they will join the cluster; (b) The vector plot of the stars' proper motion at their sky position indicates that the motions of the member stars are comparable in direction according to their membership probabilities; (c) The CMD displays member stars plotted among all stars within a 90 arcminute radius, with each member color-coded according to their membership probabilities.  }
    \label{fig:three_images_side_by_side5}
\end{figure*}

Assuming the sample includes both cluster and field stars, we compute the Mahalanobis Distance (MD) for each star and we used the three astrometric parameters, parallax and two proper motions: right ascension and declination. This distance is derived from the center of the multivariate data (the mean of each variable), which has been normalized using the covariance matrix. This procedure reduces correlations between variables and standardizes them on a single, unitless scale (\citet{(Debetal.2022)}). Thus MD transforms multi-dimensional data into one dimension (\citet{DeMaesschalcketal.2000}). In our study of the NGC 6416 cluster, we estimated the Mahalanobis Distances (MD) for 3389 stars using the parameters provided by parallax ($\pi$), proper motion in right ascension ($\mu_\alpha \cos \delta$), and proper motion in declination ($\mu_\delta$). This distribution is used as input to determine the stars' membership probability using the Gaussian Mixture Model (GMM). 

\begin{table}
    \caption{Mean values of the parameters $(\pi, \mu_\alpha \cos \delta, \mu_\delta)$ for the cluster NGC 6416 }
    \begin{tabular}{|c|c|c|c|}
        \hline
        $Radius$ & $\pi(mas)$ & $\mu_\alpha \cos \delta(mas/yr)$ & $\mu_\delta(mas/yr)$ \\ \hline
         $90$ & $0.91\pm 0.04$ & $-1.97\pm 0.10$ & $-2.32\pm 0.09$ \\ \hline
    \end{tabular}
    \label{table2}
\end{table}

A Gaussian Mixture Model (GMM) is a probabilistic technique that uses a finite number of Gaussian distributions to estimate data distribution inside a specified parameter space (\citet{McLachlan&Peel2000};\citet{Deisenroth2020}). This approach, based on Bayesian decision theory, is a popular unsupervised machine learning algorithm (\citet{McLachlan&Peel2000};\citet{pressetal.2007}). In our work, the distribution of Mahalanobis distances (MD) is effectively represented by a com-bination of two Gaussian distributions: one for cluster stars and one for field stars (\citet{(Debetal.2022)}). Let, 
$P_c(D_M \mid \mu_c, \sigma_c^2)$ and $P_f(D_M \mid \mu_f, \sigma_f^2)$ denote the Gaussian probability distributions of the cluster and the field stars, respectively. Then,

\begin{equation}
P(D_M\mid\mu,\sigma^2)=w_cP_c(D_M\mid\mu_c,\sigma_c^2) + w_fP_f(D_M\mid\mu_f,\sigma_f^2),
 \tag{4}
 \label{4}
\end{equation}
\begin{equation}
w_c + w_f =1
 \tag{5}
 \label{5}
\end{equation}

where $w_c,\mu_c,\sigma_c$ and $w_f,\mu_f,\sigma_f$ denote the weights, means, and variances of the cluster and the field stars probability distributions, respectively. The probability that a star with MD $D_{M,i}$ will belong to a class $k = (c,f)$ is determined by the responsibility, called the membership probability (\citet{Deisenroth2020}).

\begin{equation}
r_{ik}=\frac{w_k P_k(D_{M,i}\mid\mu_k,\sigma_k^2)}{w_k P_k(D_{M,i}\mid\mu_k,\sigma_k^2)},
 \tag{6}
 \label{6}
\end{equation}
Eq.\ref{4} is solved for the initial parameters ($w_c,\mu_c,\sigma_c$ and $w_f,\mu_f,\sigma_f$) using an unsupervised machine learning technique, which is termed as the expectation-maximization (EM) algorithm denoted in the literature (\citet{Dempsteretal.1977};\citet{pressetal.2007};\citet{Deisenroth2020}). The above step is applied and from the final values of ($w_k, \mu_k, \sigma_k$), the membership probability for each star belonging to the cluster 'c' (cluster membership proba-bility) is calculated using the responsibility formula,

\begin{equation}
r_{ic}=\frac{w_c P_c(D_{M,i}\mid\mu_c,\sigma_c^2)}{w_k P_k(D_{M,i}\mid\mu_k,\sigma_k^2)},
 \tag{7}
 \label{7}
\end{equation}

Stars with $r_{ic}$ > 0.50 are treated as members of the cluster ‘c’. Then GMM is used to analyze the one-dimensional MD distribution of 3389 stars. The GMM fit for the cluster, field stars, and combined distribution is shown in Fig 3(b). By setting the membership probability criterion to 0.50, 406 cluster stars are found and they belong to NGC 6416. We used Gaussian fits to these cluster members and analysed the distribution for the astrometric parameters, namely parallax and proper motions $(\pi, \mu_\alpha \cos \delta, \mu_\delta)$, so that we can obtain the mean values of them. The corresponding values of them are shown in table \ref{table2} and are also shown in Fig 4. The proper motion map in Fig. 5(a) shows the 406 cluster stars, which are color-coded according to their membership probability. Fig. 5(b) shows the appropriate motion directions $(\pi, \mu_\alpha \cos \delta, \mu_\delta)$ for these stars displayed at their locations $(\alpha,\delta)$. In Fig. 5(c), the color-magnitude diagram in the $({G}\quad{Vs}\quad{G_{BP}-G_{RP}})$) plane is shown, with each star color-coded depending on membership probability. We can observe that all of the cluster members have their motion in the same direction, implying that the cluster's stars have a strong and stable membership (\citet{(Debetal.2022)}).
 
\section{STRUCTURE PARAMETER ESTIMATION OF NGC 6416}
\label{structure}
 The radial surface density profile of an open cluster is important for understanding its spatial dis-tribution and overall structure in the $(\alpha,\delta)$ plane. This profile provides important information on the concentration of stars within the core and the steady fall toward the outskirts by measuring how the density of stars varies with radial distance from the cluster's center (\citet{(Debetal.2022)}). The profile can be fitted using a \citet{KING(1962)} fit of the following form (\citet{Carreraetal.2019};\citet{Tarricqetal.2022}),

\begin{table}
    \caption{Structural parameter values $(radius, \rho_0, r_c, r_t, \rho_{bg})$ for the cluster NGC 6416 }
    \begin{tabular}{|c|c|c|c|c|}
        \hline
        $Radius$ & $\rho_0$ & $r_c(arcmin)$ & $r_t(arcmin)$ & $\rho_{bg}$ \\ \hline
         90 & $0.26\pm0.02$ & $10.66\pm1.22$ & $60.17\pm4.06$ & $0.00\pm0.00$\\ \hline
    \end{tabular}
    \label{tab:post}
\end{table}

\begin{equation}
\rho(r) = \rho_0 \left( \frac{1}{\sqrt{1 + \left( \frac{r}{r_c} \right)^2}} - \frac{1}{\sqrt{1 + \left( \frac{r_t}{r_c} \right)^2}} \right) + \rho_{bg},
 \tag{8}
 \label{8}
\end{equation}

here $\rho_0$ represents the core density and $\rho_{bg}$ represents the background density. The terms $r_c$ and $r_t$ represent the core and the tidal radius of the cluster, respectively. Here $r_c$ is defined as the distance from the centre of the cluster (r) for which $\rho(r) = \frac{\rho_0}{2}$, and $r_t$ represents the value of r at which $\rho = \rho_{bg}$. To acquire the cluster's radial density profile we calculate the radial distance of the $i^{th}$ cluster member $(\alpha_i,\delta_i)$ from the the center of the cluster $(\alpha_0,\delta_0)$ and is calculated by the equation,
\begin{equation}
\cos r = \cos \delta_i \cos \delta_0 \cos(\alpha_i - \alpha_0) +  \sin \delta_i \sin \delta_0, 
 \tag{9}
 \label{9}
\end{equation}

here i goes from i=1,2,3,…….,N. N denotes the number of cluster members that we calculated at the previous steps. We also need to convert the values of $(\alpha,\delta)$ into arcmin from radians. The formula for calculating stellar surface density $(\rho_i)$ is: $\rho_i = \frac{N_i}{A_i}$ where $N_i$ denotes the number of stars within the $i^{th}$ ring, bordered by the inner radius $r_i$ and the outer radius $r_{i+1}$. The size of the ring is provided by $A_i = \pi (r_{i+1}^2 - r_i^2)$, and $\rho_i$ denotes the number of stars per square arcminute within each ring. Similarly the uncertainty in density for each ring, assuming Poisson statistics, is given by $\sigma_{pi} = \frac{\sqrt{N_i}}{A_i}$ (\citet{(Debetal.2022)}).

\begin{figure}
    \centering
    \includegraphics[width=1\linewidth]{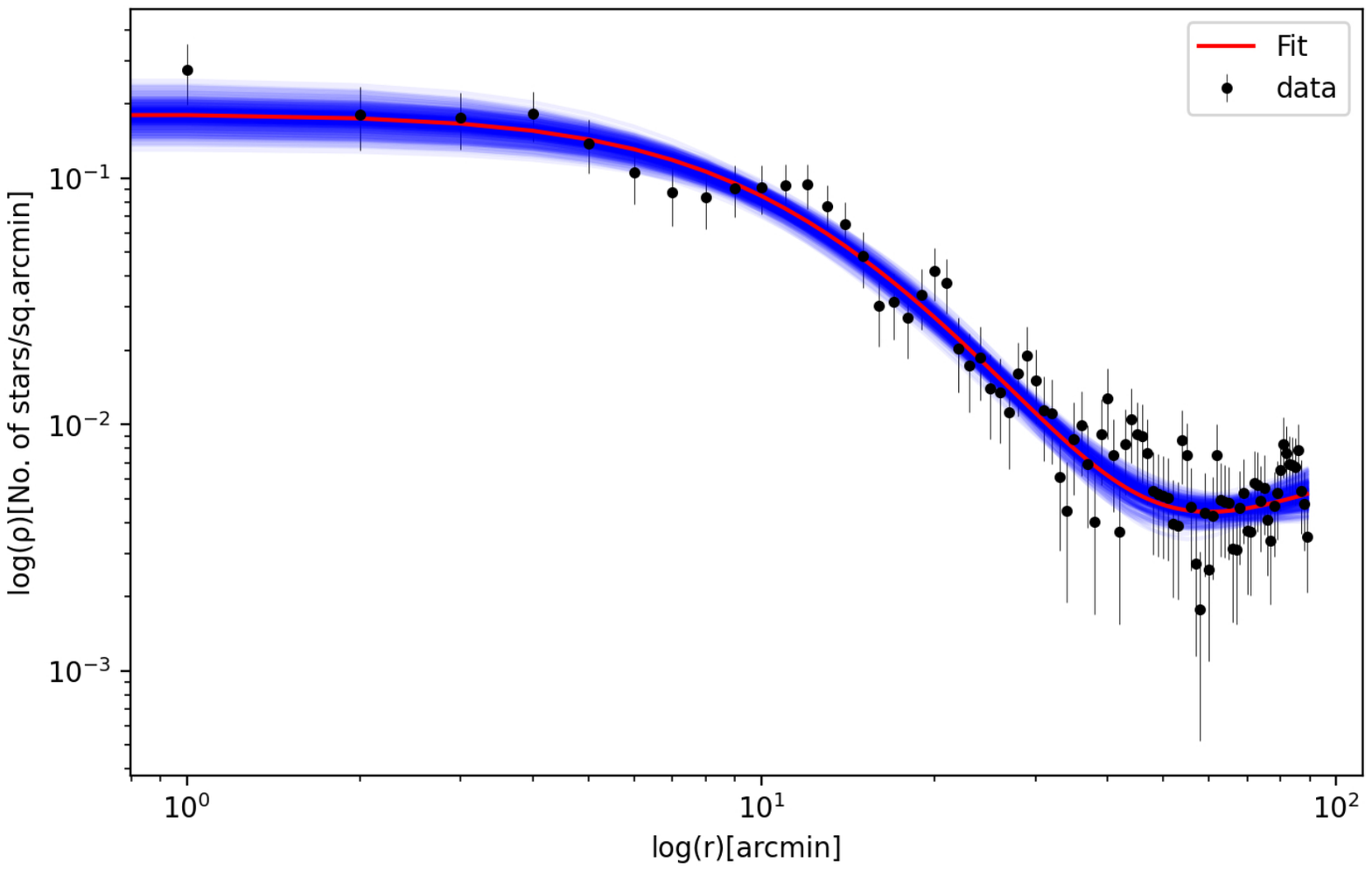}
    \caption{Plot displays the radial density profile fit findings for cluster members of NGC 6416 within a 90 arcmin radius. The uncertainty range, represented by the error bars, is $\pm1\sigma$ Poissonian. The MCMC chains generated three hundred random samples and blue lines represent random sample fits.}
    \label{fig:enter-label}
\end{figure}

\begin{figure}
    \centering
    \includegraphics[width=1\linewidth]{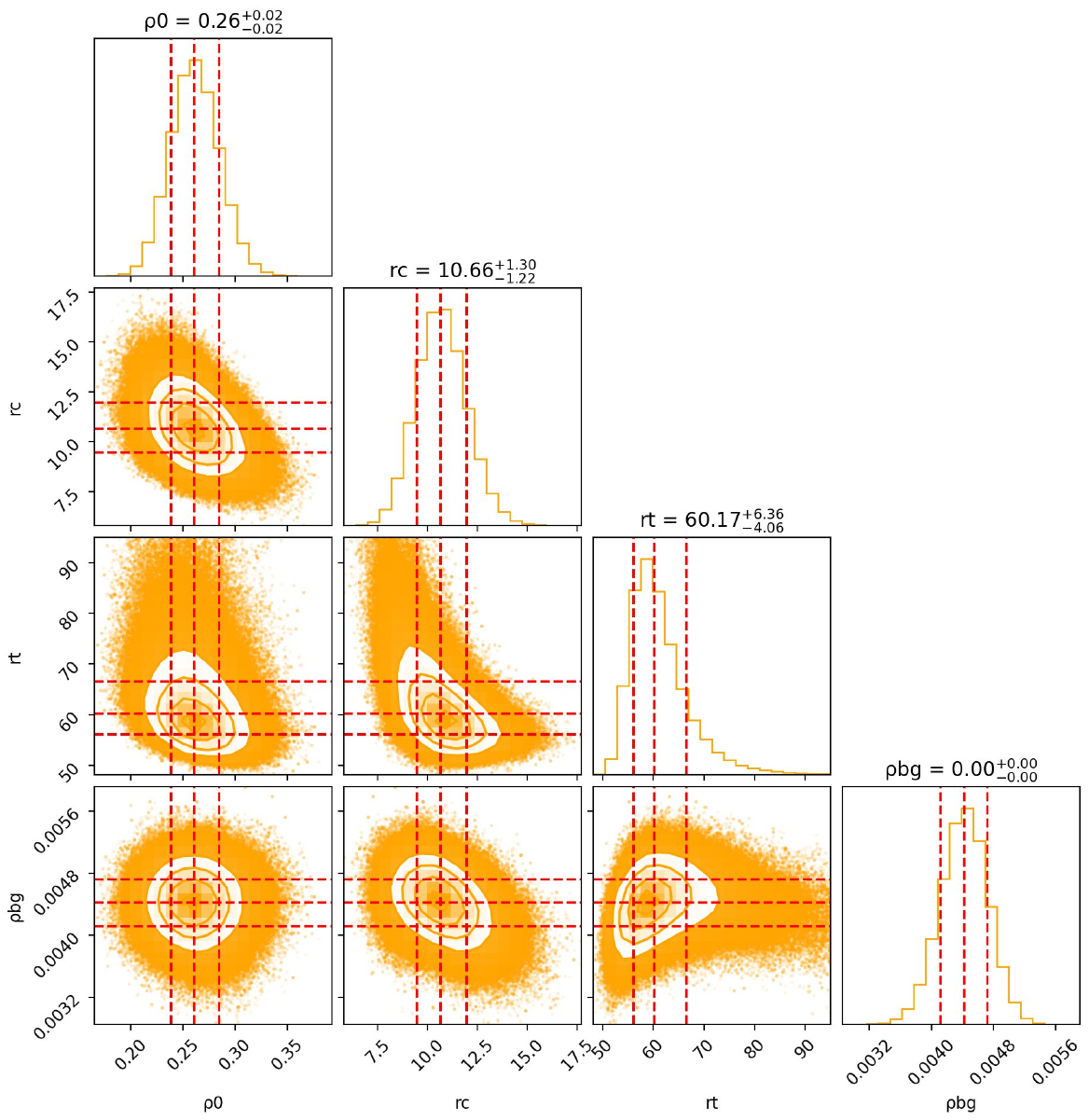}
    \caption{The MCMC study of the King profile fit for the cluster NGC 6416 yielded the marginalized 
posterior distribution and uncertainties of the model parameters, taking into account member sources up to 90 
arcmin. The sample mean is $50^{th}$ and the $16^{th}$ and $84^{th}$ percentile differences (up and down) are represented by the statistical uncertainty. On each histogram, the $16^{th}$, $50^{th}$, and $84^{th}$ percentiles are indicated by the magenta dashed vertical lines, respectively}
    \label{fig:enter-label2}
\end{figure}

The radial density profile was fitted with Equation \ref{8}, taking into account all final cluster members discovered within the chosen search radius(i,e, within 90 arcmin). The background density $(\rho_{bg})$ should be zero if there is no contamination from field stars in the membership analysis. Still a modest non-zero value of $\rho_{bg}$ shows the extent of contamination influencing the cluster, providing an estimate of background star infiltration (\citet{Tarricqetal.2022}). The parameters and their uncertainties are calculated using the Bayesian Markov Chain Monte Carlo (MCMC) technique, which is implemented in Python using the emcee package (\citet{Foreman-Mackey_etal.2013}). Fig 7 shows the marginalized posterior distributions for each parameter as generated from the MCMC chains. To ensure convergence and adequate parameter exploration, the sampling procedure uses 2000 walkers, 1000 iterations, and 200 burn-in steps \citet{(Debetal.2022)}.

The high symmetry of the distribution implies that the mean and median are about equal. The parameters' best-fitting value is the $50^{th}$ percentile of the distribution, which is the mean. The statistical uncertainties are defined as the range between the $16^{th}$ and $84^{th}$ percentiles, which provide lower and upper bounds, respectively. Figure 6 shows the observed data, which is represented by black dots. The best-fitting line is shown in red, while 300 random sample fits from the MCMC chains are represented by blue lines. These blue lines illustrate the range of uncertainty in the model parameters. The  best fitted parameter values that we found is shown in the table \ref{tab:post}.

\section{Isochrone Properties OF NGC 6416}
\label{isochrone}
A comprehensive analysis of these astronomical ensembles necessitates an understanding of the fundamental attributes of star clusters, including distance, metallicity, age, extinction or reddening, binary fraction, and the total-to-selective extinction ratio. Each piece of data provides significant insights into the formation, evolution, and overall galaxy-wide context of the cluster (\citet{(SBiswasetal.2024)}). In smaller clusters, trigonometric parallax is frequently used to determine distance; for larger clusters, variable stars and main sequence fitting are employed. The specific distances are necessary for generating physical scales and converting apparent magnitudes to absolute magnitudes. Spectroscopy, which examines absorption bands in stellar spectra, is used to measure metallicity, which influences stellar evolution. This parameter impacts cluster CMDs and evolutionary models by determining a star's color and temperature. To ascertain the age of the cluster, isochrone fitting involves overlaying theoretical isochrones on the CMD. The CMD's turn-off point, which indicates when stars transition from the main sequence to the giant branch, provides a clear indication of the cluster's age.

Extinction change accounts for the effects of interstellar dust by adjusting the measured hues and magnitudes. While infrared observations can mitigate some dust effects, color excess quantifies this by comparing observed colors to intrinsic colors. Binary or multi-star systems are prevalent in clusters. Identifying and defining these systems is crucial for conducting accurate mass and dynamical research, typically achieved through photometric variability or radial velocity observations. The Total-to-Selective Extinction Ratio serves as a valuable tool for explaining how dust scatters and absorbs light, which is vital for accurately adjusting measured colors and magnitudes. It is often determined using spectral analysis or multi-band photometry and varies with the size and composition of dust grains ( \citet{Bressan2012};\citet{Girardi2002}).

Based on the stellar evolution code from the Modules for Experiments in Stellar Astrophysics (MESA), the MESA Isochrones and Stellar Tracks (MIST) is a powerful tool that we applied to the data. The MIST isochrone serves as a theoretical model that predicts the positions of stars on a CMD, taking into account variables such as age, metallicity, and distance. As illustrated in Figure 8, we fitted the observed CMD to the MIST isochrone using the best-fitting parameters.

\begin{figure}
    \centering
    \includegraphics[width=1\linewidth]{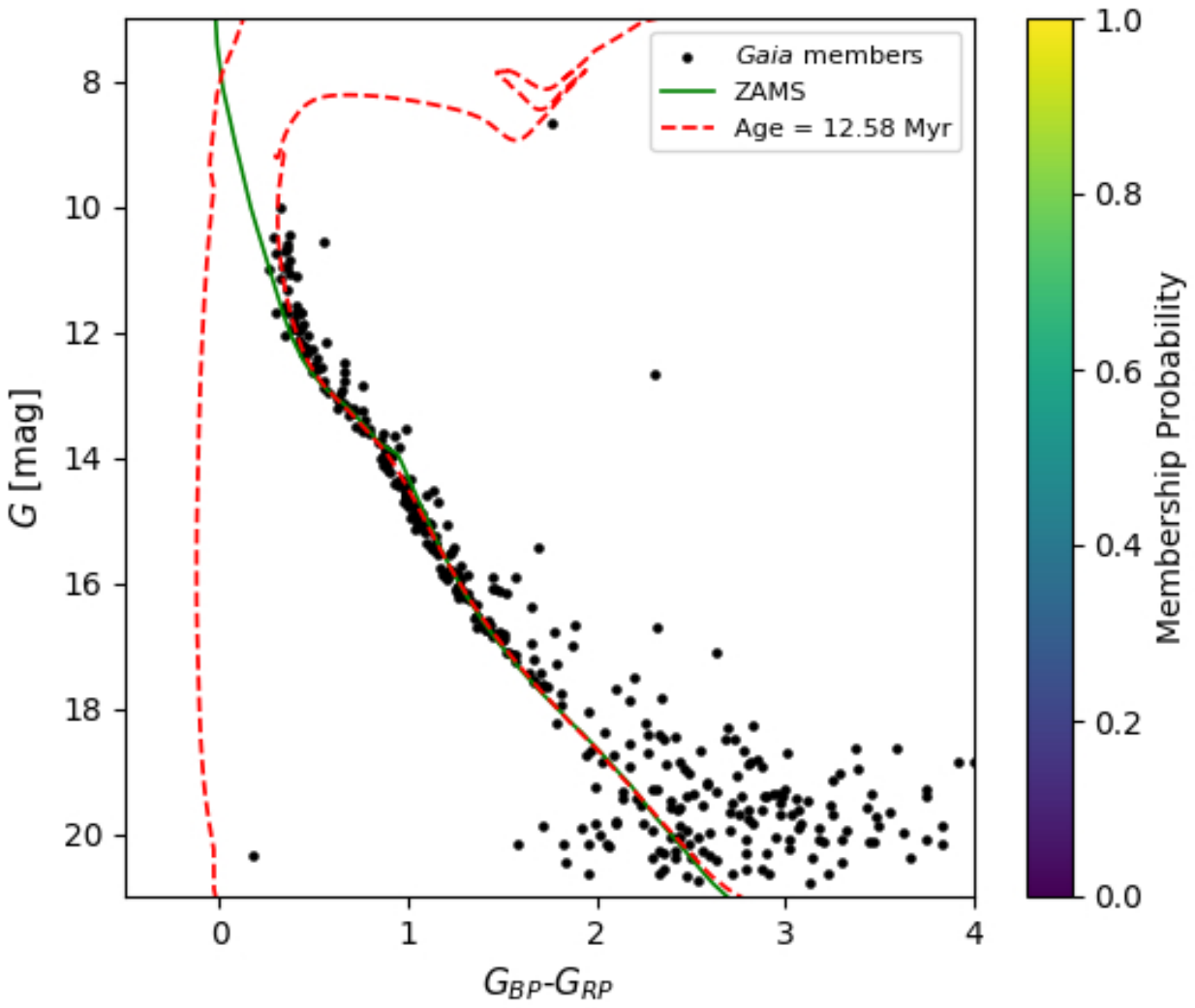}
    \caption{Color-magnitude diagram for NGC 6416 cluster members with membership probabilities exceeding 0.50. The ZAMS (Zero Age Main Sequence) is represented by the solid green line, plotted from MIST isochrones. The PMS isochrone for an age of 12.58 Myr from the MIST isochrones is depicted by the colored dotted lines.}
    \label{fig:enter-label}
\end{figure}

Figure 8 depicts the best-fitted isochrone produced. The best-fit values obtained are  metallicity (z) = $0.032 \pm 0.0015$, log(age) = $8.1 \pm 0.012$, binary fraction ($b_{frac}$) = $0.419 \pm 0.021$, visual extinction ($A_V$) = $0.995 \pm 0.058$ mag, total-to-selective extinction ratio ($R_V$) = $3.064 \pm 0.102$, and distance modulus ($(m - M)_O$) = $10.13 \pm 0.01$ and we estimated distance of the cluster 
NGC 6416 is found to be 1021 pc. Using the equation $E(B - V)$
= $A_V /R_V$, the reddening $E(B - V)$
 is estimated to be approximately $0.32 \pm 0.02$ mag.

\section{ORBITAL PARAMETERS OF NGC 6416}
\label{orbital}
Determining the birth radii and galactic distribution of open clusters requires an understanding of their kinematics and dynamics. This information sheds light on the development, evolution, and general role of these stellar groups in the process of star formation within the galaxy (\citet{Sheikh.A.H&MedhiBimanJ2024}). We may study the geographical distribution of open clusters in the solar neighborhood and gain insight into their origins and current positions on the Galactic plane by measuring important parameters such as radial velocities, distances, and ages (\citet{Yontan2023}). NGC 6416's orbit was examined using the MWPotential2014 model from the Galactic dynamics library, galpy (\citet{Bovy2015}). This model proposes an axisymmetric potential for the Milky Way, incorporating three essential elements of the Galactic potential: a huge, spherical dark-matter halo based on \citet{Navarroetal.(1996)}, a spherical bulge described by \citet{Bovy2015}, and a Galactic disk outlined by \citet{Miyamoto.M&Nagai.R.1975}.

\begin{figure}
    \centering
    \includegraphics[width=1\linewidth]{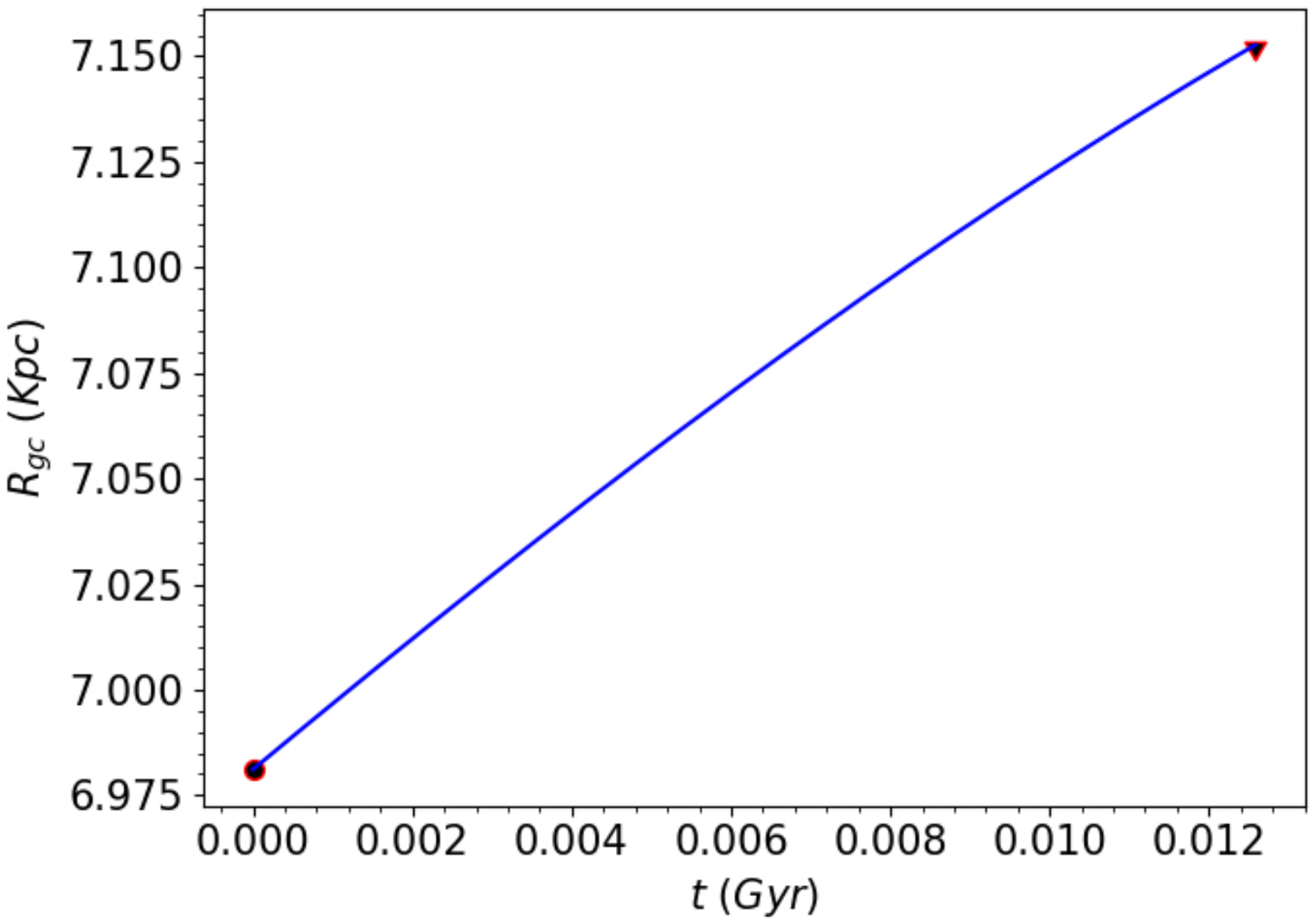}
    \caption{The birth radii and Galactic orbits of NGC 6416 in the $R_{gc} \times t$ planes are illustrated. The red triangle indicates the current location of NGC 6416, while the red arrow represents its velocity vector.}
    \label{fig:enter-label}
\end{figure}

To complete the orbit integration procedure, the radial velocity parameter is needed. We utilized the mean radial velocity $V_{\gamma}=-27.50 \pm 9.94$ km/s taken from \citet{LoktinA.V.2017}, for NGC 6416. The orbit integration was based on the mean proper motion components $\mu_\alpha \cos \delta = -1.97 \pm 0.10$ $masyr^{-1}$ and $\mu_\delta = -2.32 \pm 0.09$ $masyr^{-1}$, which we found from section \ref{member}, the isochrone distance, $d_{iso} = 1021 pc$ from section \ref{isochrone}, as well as the central equatorial coordinates ($\alpha= 266.014$  , $\delta= -32.344$) from \citet{Tarricq.Y.2021}. We used the values of $R_{gc}$ = 8 kpc and $R_{rot}$ = 220 $kms^{-1}$ for the Sun's orbital velocity and galactocentric distance, respectively (\citet{Bovy2015};\citet{Bovy2012}). We also used 25 kpc for the Sun's distance from the Galactic plane (\citet{JuriÄ2008}).

\begin{table}
    \caption{Orbital parameter values $(R_a, R_p, Z_{max}, P_{orb})$ for the cluster NGC 6416 }
    \begin{tabular}{|c|c|c|c|}
        \hline
         $ R_a(kpc)$ & $R_p(kpc)$ & $Z_{max}(kpc)$ & $P_{orb}(Myr)$ \\ \hline
          $7.152\pm0.42$ & $6.981\pm0.028$ &$0.103\pm0.21$ & $140\pm4$\\ \hline
    \end{tabular}
    \label{tab:orbital}
\end{table}

We have carried out a forward integration of the cluster's orbit with a step size of 1 Myr across a time span of up to 13 Gyr in order to estimate its likely current position. The cluster's "side view" on the Galactic center and plane is depicted in Figure 9 along with its distance from both. The symbol $Z$ indicates the vertical dis-tance from the Galactic plane, whereas $R$ $(gc)$ indicates the Milky Way's Galactocentric distance. We performed orbit calculations for an earlier epoch that corresponds to the age of the cluster, which is 12.58 Myr, in order to estimate the likely birth radius of NGC 6416 was found to be $R(kpc)$ = $6.96\pm0.01$. The cluster's distance on the $R_{gc} \times t$ plane is shown in Fig. 10 with time, where t is the cluster's age (\citet{Yontan2023}), the eccentricity was found as $e$ = $0.0121\pm0.0098$. Orbit integration gives the following parameters of the cluster NGC 6416 and is shown in the table \ref{tab:orbital} .

\begin{figure}
    \centering
    \includegraphics[width=1\linewidth]{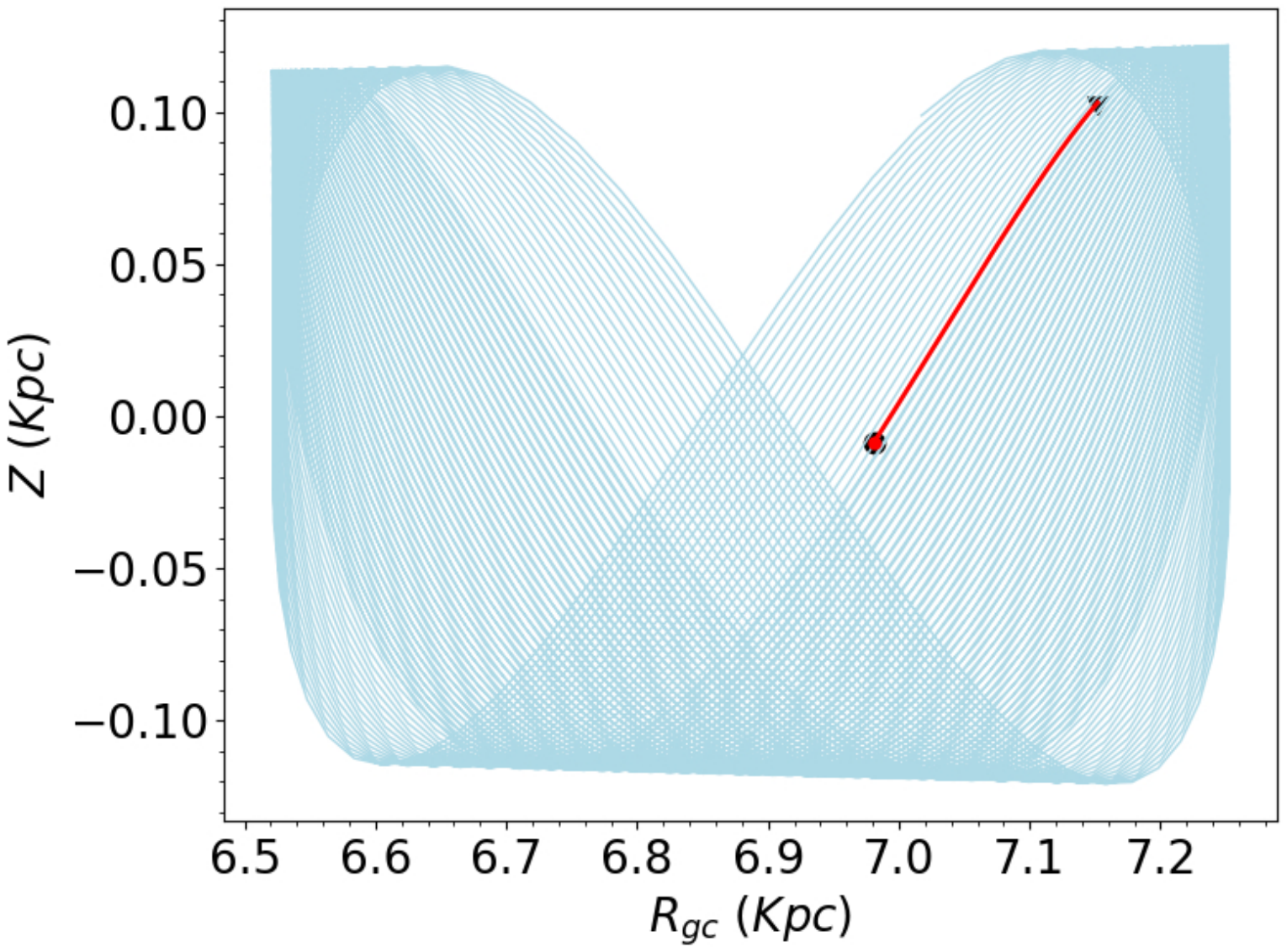}
    \caption{The NGC 6416 Galactic orbits and birth radii in the $R_{gc} \times z$ planes. The birth places are shown by the black circle, and NGC 6416's current location is shown by the black triangle.}
    \label{fig:enter-label}
\end{figure}

\section{Results}
\label{results}
We determined the age, metallicity, and distance of the open star cluster NGC 6416 using photometry and astrometry from GAIA EDR3. The members of the star cluster NGC 6416 were identified using the GAIA astrometry data $(\pi, \mu_\alpha \cos \delta, \mu_\delta)$. To ascertain the membership probability of the open cluster NGC 6416, we additionally employed an unsupervised ensemble machine learning technique \citet{(Debetal.2022)}. Upon identifying the cluster members, we calculated the radial density profiles of the open cluster NGC 6416 and determined the king fit parameters of the cluster $(\rho_0, r_c, r_t, \rho_{bg})$.

Within a 90 arcmin search radius, we identified 406 stars with a membership probability > 0.50 by employing an unsupervised ensemble machine learning approach on the NGC 6416 cluster. Utilizing a Gaussian fit to the distribution, the mean values of the parameters are $\pi(mas)$ = $0.91\pm 0.04$, $\mu_\alpha \cos \delta(mas/yr)$ = $-1.97\pm 0.10$, $\mu_\delta(mas/yr)$ = $-2.32\pm 0.09$, which is almost close to what was found on \citet{Cantat_Gaudin_2020}. The vector map, illustrating that nearly all cluster members are moving in the same direction, demonstrates the robustness of the star cluster membership determination. The Bayesian Markov Chain Monte Carlo (MCMC) method, along with the Python EMCEE module \citet{Foreman-Mackey_etal.2013}, is employed to ascertain the values and uncertainties of the King fit parameters \citet{KING(1962)}. They are written as follows $\rho_0$ = $0.26\pm0.02$, $r_c(arcmin)$ = $10.66\pm1.22$, $r_t(arcmin)$ = $60.17\pm4.06$, $\rho_{bg}$ = $0.00\pm0.00$. Figure 7 displays the marginalized posterior distribution and model parameter 
uncertainties of the King profile fit for the cluster NGC 6416.

The fundamental characteristics of the cluster NGC 6416 were determined as: The total 
metallicity, z = $0.032\pm0.0015$, we also estimated the age of the cluster from the isochrones \citet{PerrenG.I.} is to be log(age) = $8.1 \pm 0.012$, which is almost equivalent to what they \citet{Ray_2022} calculated for NGC 6416 as log(age)= 8.36. The binary fraction or multiple star systems is found to be binary fraction ($b_{frac}$) = $0.419 \pm 0.021$, total-to-selective extinction ratio ($R_V$) = $3.064 \pm 0.102$, visual extinction ($A_V$) = $0.995 \pm 0.058$ mag, the reddening is estimated to be approximately $E(B - V)$ = $0.32 \pm 0.02$ mag,the distance modulus is found to be ($(m - M)_O$) = $10.13 \pm 0.01$, and the distance of the cluster NGC 6416 is found to be $1021\pm0.03$ pc, which almost close to 1020 pc (\citet{E.Poggio2021}).

Further the analysis of the of galactic orbits has revealed certain parameters for the open star cluster 
NGC 6416 and they are given as follows $R_a(kpc)$ = $7.152\pm0.42$, $R_p(kpc)$ = $6.981\pm0.028$, $e$ = $0.0121\pm0.0098$, $Z_{max}(kpc)$ = $0.103\pm0.21$, $P_{orb}(Myr)$ = $140\pm4$. The birth radius was found to be $R(kpc)$ = $6.96\pm0.01$, here the perigalactic and apogalactic distances shows that the orbit is inside the solar circle, indicating that it has a higher metallicity compared to clusters located farther out (\citet{Sheikh.A.H&MedhiBimanJ2024}).

\section{summary and conclusion}
\label{summary}
NGC 6416 is a notable open star cluster whose properties have been thoroughly studied using data from Gaia EDR3 astrometry and statistically. To correctly identify cluster members, we 
used Gaia's precise astrometric measurements, including parallax, PmRa, and PmDec. According to our analysis, the parallax and proper motion distributions show that cluster 
members are grouped around a mean value. All stars travel in the same direction, according to a velocity vector diagram that depicts their proper movements at their individual sky positions. A neat diagram is also produced when a color-magnitude diagram is made with the cluster members. The method used in this study produces very reliable results for the cluster membership analysis. In a one-dimensional MD distribution, we found two bimodal peaks: one for field stars and one for cluster stars. These fit well with two Gaussian distributions: the field having a wider peak and the cluster having a sharper peak. The clustering method established 
in this work only applies to open star clusters \citet{(Debetal.2022)}. 

From the density distribution of the cluster using king fit \citet{KING(1962)}, we found that a low central density is observed, as we know a high central density indicates a densely packed core, which is commonly observed in 
older or more actively evolving clusters, since our star cluster NGC 6416 does not have a high central density it is found that it is not an evolved star cluster. The core radius reflects the size of the cluster's densest region and is essential for understanding the internal dynamics and 
potential interactions between stars. The tidal radius indicates the cluster's effect and is affected 
by the galaxy's gravitational pull. Larger tidal radii indicate a more gravitationally bound 
cluster, in our case the open cluster NGC 6416 is fairly bounded by galaxy’s gravitational pull (\citet{2008gady.book.....B}.

The metallicity number indicates a non-solar-like chemical composition, implying that the cluster has not much of a similar elemental abundance to the Sun, as is common for clusters located a bit above the MilkyWay's disk, as it has a significantly larger vertical dis-
tance from the Galactic plane about $Z_{max}(kpc)$ = $0.103\pm0.21$, it shows that NGC 6416 is an intermediate-age open star cluster with non-solar-like metallicity, a significant binary star percentage, and mild extinction and reddening properties. Its age and distance make it a fascinating target for future research into star formation, cluster dynamics, and stellar evolution. The isochrone fitting, give a solid foundation for understanding its physical and chemical features, adding vital data to the larger study of galactic star clusters (\citet{1995ARA&A}).

We also found that the isochrones from Fig 8 show a significantly low number of YSOs (Young Stellar Objects). YSOs are stars in the early stages of their evolution, typically located in star-forming regions within molecular clouds. Thus, NGC 6416 suggests that most YSOs have matured into main-sequence stars, leading to a reduced current population of YSOs. This may also indicate a scarcity of YSOs, implying limited or no recent star formation activity within the cluster. Possible reasons for this could include the depletion of molecular gas necessary for star formation or environmental factors that inhibit such processes, such as low gas density or disruptive gravitational interactions.




\bibliographystyle{mnras}
\bibliography{main} 




\bsp	
\label{lastpage}
\end{document}